\documentclass[%
    reprint,
    showpacs,
    amsmath,amssymb,
    longbibliography,
    aps,
    pra,
    floatfix,
    superscriptaddress
]{revtex4-2}
\usepackage{xcolor}
\usepackage{soul}
\usepackage[utf8]{inputenc}	
\usepackage[T1]{fontenc}	
\usepackage[]{graphicx}		
\usepackage{xcolor}
\usepackage{gensymb}
\usepackage{amsmath, mathdots}
\usepackage{physics}
\usepackage{siunitx}
\usepackage{comment}
\usepackage[english]{babel}
\usepackage{color}
\usepackage{bm}
\usepackage[colorlinks]{hyperref}
\hypersetup{
    colorlinks=true,
    linkcolor=blue,
    filecolor=blue,      
    urlcolor=blue,
    citecolor=blue
}

\renewcommand{\a}{\hat{a}}
\renewcommand{\b}{\hat{b}}

\newcommand{\iu}{\text{i}}
\newcommand{\T}{\text{T}}
\newcommand{\hc}{\text{H.c.}}
\newcommand{\cov}{\text{cov}}
\DeclareMathOperator{\sinc}{sinc}
\DeclareMathOperator{\sign}{sign}
\DeclareMathOperator{\diag}{diag}
\renewcommand{\Re}{\text{Re}}
\renewcommand{\Im}{\text{Im}}

\newcommand{\inprod}[2]{{\langle #1, #2 \rangle}}

\usepackage{mathtools}

\begin{document}

\title{Phase-space open-systems dynamics of second-order nonlinear interactions with pulsed quantum light.}

\author{Emanuel Hubenschmid}
\email[]{emanuel.hubenschmid@uni-konstanz.de}
\affiliation{Department of Physics, University of Konstanz, D-78457 Konstanz, Germany}

\author{Victor Rueskov Christiansen}
\email[]{victorrc@phys.au.dk}
\affiliation{Department of Physics and Astronomy, Aarhus University, Ny Munkegade 120, DK-8000 Aarhus C, Denmark}

\begin{abstract}
\noindent
	The theoretical description of broadband, multimode quantum pulses undergoing a second-order $\chi^{(2)}$-nonlinear interaction can be quite intricate, due to the large dimensionality of the underlying phase space.
    However, in many cases only a few broadband (temporal) modes are relevant before and after the nonlinear interaction.
	Here we present an efficient framework to calculate the relation between the quantum states at the input and output of a nonlinear element in their respective relevant modes.
	Since the number of relevant input and output modes may differ, resulting in an open quantum system, we introduce the generalized Bloch-Messiah decomposition (GBMD), reducing the description to an equal number of input and output modes.
    The GBMD enables us to calculate the multimode Wigner function of the output state by convolving the rescaled Wigner function of the reduced input quantum pulse with a multivariate Gaussian phase-space function.
    We expand on this result by considering two examples input states: A Fock state in a single broadband mode and a two-mode squeezed vacuum, both in the THz-frequency regime, up-converted to a single output broadband mode of optical frequencies.
	We investigate the effect, the convolution and thermalization due to entanglement breakage have on the output Wigner function by calculating  the von Neumann entropy of the output Wigner function.
    The methods presented here can be used to optimize the amplification or frequency conversion of broadband quantum states, opening an avenue to the generation and characterization of optical quantum states on ultrafast time scales.
\end{abstract}
\maketitle

\noindent
\section{Introduction}
Nonlinear optics has been essential for the development of quantum optics.
The generation of quantum pulses of light, such as the squeezed vacuum, enables a plethora of quantum information technologies spanning from quantum metrology (including the subshotnoise detection of gravitational waves \cite{Vahlbruch2010,Aasi2013} or precise spectroscopy \cite{Dorfman2016,Mukamel2020,Datta2025,Khan2025}), quantum key distribution \cite{QKD} and quantum networks \cite{kimble2008quantum}, to the generation of even more resourceful states like photon subtracted squeezed states \cite{Walschaers2017}.

\begin{figure}
	\centering
	\includegraphics[width=0.45\textwidth]{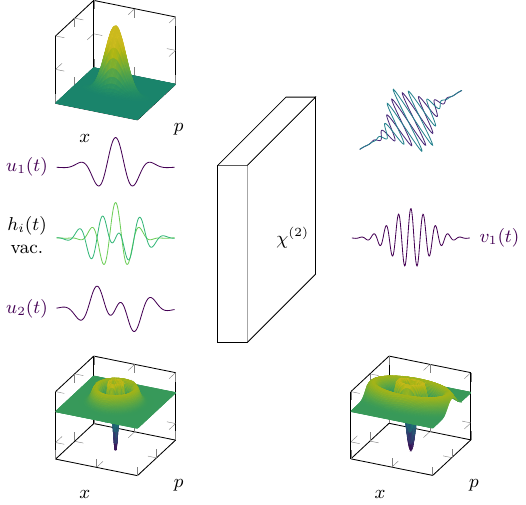}
	\caption{
		A schematic representation of the scenario considered in this work.
		In the most general setting, a quantum pulse of light, composed of multiple temporal modes with profiles $u_i(t)$, traverses a nonlinear medium with a second-order nonlinear susceptibility $\chi^{(2)}$.
		After the nonlinear interaction some, possibly different, modes $v_i(t)$ are relevant, e.g., for detection.
		An example of such a setting is electro-optic sampling or nonlinear homodyne detection.
		Since the set of input and output modes is not necessarily closed under the transformation by the nonlinear interaction, some additional modes $h_i(t)$ in the ground state (vacuum) are required for the description.
		The main achievement of this work is a relation between the input quantum pulse and the quantum state occupying the output modes based on the (multimode) Wigner function in Eq.~\eqref{eq:outputWignerfunction1}.
		This relation only includes the modes $u_i(t)$ and $v_i(t)$, but not $h_i(t)$, which we achieve by introducing the generalized Bloch-Messiah decomposition (GBMD) in Sec.~\ref{sec:gbmd}.
	}
	\label{fig:nl_interaction}
\end{figure}

In recent years, the nonlinear interaction of light has enabled the generation of broadband quantum pulses \cite{Riek2017,Kawasaki2024,Spasibko2017} and the ultrafast detection of (quantum) light \cite{Riek2015, Riek2017}, a development which requires novel theoretical tools to describe the intricate mode structure and statistics of broadband quantum states.
While the (second-order) nonlinear interaction of a few monochromatic modes of light can be described using symplectic matrices \cite{Adesso2014,Weedbrook2012}, it is not always possible to extend this formalism to pulsed quantum states.
After parametric amplification, the phase-space dimension of the broadband output pulses is in general twice the dimension of the input phase space \cite{parametric-amplification-quantum-pulse}, and can possibly be infinite, for example, if the input is in the ground state (vacuum) \cite{Wasilewski2006, parametric-amplification-quantum-pulse}.
However, in many scenarios, only a few or even a single pulsed mode is relevant for further processing, e.g., if the quantum pulse is detected after the nonlinear interaction \cite{Hubenschmid2022}, which results in an open-system dynamics.
Thus, for many applications, the input-output relation of the nonlinear interaction may be between a different number of relevant input and output modes, as schematically sketched in Fig.~\ref{fig:nl_interaction}.

Here, we present a method based on the symplectic singular value decomposition (SVD) introduced by Xu \cite{Xu2003,Xu2005}, which enables a reduction of the number of input or output modes to a set of effective modes with matching number of input-output modes.
Since the method is based on a symplectic version of the SVD we refer to it as the generalized Bloch-Messiah decomposition (GBMD).
While the usual Bloch-Messiah decomposition is a remarkable, resourceful tool in nonlinear quantum optics, as long as the dynamics is unitary and closed, \cite{Houde2024a,Wasilewski2006, Christ2013, Ansari2018}, the generalized version presented here is applicable to arbitrary sets of input and output modes, which comprise the open systems.
The effect of a mismatch between the transformed input modes and the relevant output modes on the quantum state is a non-unitary evolution during the effective nonlinear interaction.
Thus, the output quantum state can be a mixed state, even if the quantum state at the input of the nonlinear interaction is pure.
We describe this effect using Wigner function representation of the quantum state in phase space.

The main result of this work is the phase-space input-output relation of the quantum states in Eq.~\eqref{eq:outputWignerfunction1} connecting the state prior and after the nonlinear interaction.
The Wigner function representing the output quantum state is given by a convolution of the Wigner function corresponding to transformed and reduced input state with a Gaussian phase-space function and a subsequent rescaling.

As an example of this result, for which a different number of relevant input and output pulses is encountered, we consider the ultrafast detection of quantum light, e.g., using electro-optic sampling (EOS) to access the terahertz (THz) or mid-infrared (MIR) frequency range.
Direct photo detection in the THz frequency range is challenging, since the THz-photon energy is too low for efficient photo detection, but the dynamics is too fast for electronics.
Electro-optic sampling utilizes a nonlinear crystal to up-convert the light to optical frequencies.
The interaction is gated by an ultrashort classical coherent pulse.
After up-conversion, the optical frequency component of the quantum state in the mode of the classical pulse is measured using homodyne detection, thus corresponding to the scenario of only one relevant output mode.
By time-delaying the classical and low frequency pulse, the dynamics of the field in the THz/MIR range can be scanned in the time domain \cite{Leitenstorfer1999,Gallot1999,Sulzer2020,Moskalenko2015,Riek2015,Riek2017,Kempf2024,Guedes2019,Kizmann2019,BeneaChelmus2019,Lindel2020,Lindel2021,Virally2021,Beckh2021,Onoe2022,Kizmann2022,Guedes2023,Settembrini2022,Settembrini2023,Lindel2023,Yang2023,Lindel2024,BeneaChelmus2025,Yang2025,Adamou2025}.
While electro-optic sampling has already been applied to the detection of the vacuum \cite{Riek2015,BeneaChelmus2019} and squeezed vacuum pulses \cite{Riek2017} in the THz regime, recent proposals extend this technique to an ultrafast quantum state tomography of single mode pulsed quantum states \cite{Hubenschmid2024,Yang2023,Onoe2023} or multimode Gaussian quantum states \cite{Hubenschmid2025,Yang2025}.
Another case, related to electro-optic sampling, is nonlinear photo-detection, which utilizes a nonlinear interaction prior to photo-detection to amplify one quadrature, thus enabling the measurement of that quadrature \cite{Kalash2023,Kalash2025}, or fast photo detection \cite{Yanagimoto2023,Sendonaris2024}.

Here, we describe the up-conversion using the formalism of the first-order unitary introduced in \cite{Onoe2022,Hubenschmid2024}, which can be understood as a renormalized, first-order perturbation theory based on the Baker–Campbell–Hausdorff formula. We consider two examples of the phase-space input-output relation.
In the first example, a single pulsed mode Fock state in the THz range is up-converted to a single broadband output mode in the optical frequencies.
We show theoretically that the up-conversion exhibits two different regimes, which we call the squeezing and beam-splitting regime, determined by the central frequency of the output mode.
The regime drastically influences the dependence of the output quantum state on the strength of the nonlinear interaction.
Additionally, we show how to optimize the up-conversion in the beam splitting regime.
The second example consists of a two-mode squeezed vacuum pulse in the THz range up-converted to a single broadband output mode in the optical frequencies.
In contrast to the single input mode, applying the generalized Bloch-Messiah decomposition to reduce the phase-space dimension of the input state can lead to thermalization originating from the breakage of entanglement present in the two-mode squeezed vacuum.
We quantify the thermalization using the von Neumann entropy and show how it depends on the central frequency of the output mode.
In the example considered here, the thermalization in the squeezing regime originates from state-independent entanglement generated in the up-conversion.
In the beam-splitter regime, the entanglement present in the quantum state prior to the up-conversion becomes relevant, which could enable the direct measurement of entanglement in a two-mode squeezed THz pulse.

This work is organized as follows.
In Sec.~\ref{sec:broadbandModes} we give a short overview of the formalism used in this work, including broadband modes and the multimode Wigner function.
In Sec.~\ref{sec:in_out_op}, the input-output relation of the quadrature operators is presented.
The generalized Bloch-Messiah decomposition is introduced in Sec.~\ref{sec:gbmd} which is then utilized in Sec.~\ref{sec:in_out_ps} to derive the phase space input-output relation of the pulsed quantum state.
Sec.~\ref{sec:first_order_unitary} elaborates on the first-order unitary approach used to calculate the operator input-output relation for the THz-to-optical frequency up-conversion.
Sec.~\ref{sec:examples} builds upon the theory derived in the previous sections using the example of a Fock state in a single broadband THz mode and the two-mode squeezed vacuum in two broadband THz modes, both up-converted to a single broadband mode in the optical frequencies.
The work is summarized and concluded in Sec.~\ref{sec:conclusion} with an outlook on further research.

\section{A brief introduction to broadband modes}\label{sec:broadbandModes}
Quantum optics is based on the assumption that light of a single color with (angular) frequency $\omega$ consists of quanta of a well defined energy, which are annihilated or created by the operators $\hat{a}(\omega)$ or $\hat{a}^\dagger(\omega)$ (as shorthand $\hat{a}_\omega$ or $\hat{a}^\dagger_\omega$).
This single frequency basis may be good for some setups, but since nonlinear optics enables the interaction of light in an ultrabroad band of frequencies, the description of all individual monochromatic modes can be computationally exhaustive. In this case, it can be instructive to define broadband (temporal) modes which comprise of a continuum of frequencies instead of a monochromatic wave \cite{Brecht2015,Raymer2020}.
The quanta of a broadband mode, with frequency distribution $u(\omega)$, are annihilated by $\hat{a}_{u} = \int_0^\infty u^\ast(\omega) \hat{a}_\omega \dd \omega$.
The extension to $N$ broadband modes $u_j(\omega)$, requires the mode functions to be orthogonal with respect to the Hilbert-Schmidt inner product $\inprod{u_i}{u_j} = \int_0^\infty u_i^\ast(\omega) u_j(\omega) \dd \omega$, to ensure the canonical commutation relations are fulfilled, $[\hat{a}_{u_i}, \hat{a}_{u_j}^\dagger] = \delta_{ij}$.
In this case, the modes are said to be orthogonal.

The mode operators can be used to define quadratures $\hat{x}_{u_j} = \frac{1}{\sqrt{2}}(\hat{a}_{u_j} + \hat{a}_{u_j}^\dagger)$ and $\hat{p}_{u_j} = \frac{1}{\sqrt{2}\iu}(\hat{a}_{u_j} - \hat{a}_{u_j}^\dagger)$, as well as a quantum phase space $\hat{\Vec{\Gamma}} = (\hat{x}_{u_1}, \ldots, \hat{x }_{u_N}, \hat{p}_{u_1}, \ldots, \hat{p}_{u_N})^\T$, where the transpose of an operator valued vector $\hat{\vec{v}}$, denoted by $\hat{\vec{v}}^\T$, is only applied to the vector not the elements, i.e., the quadrature operators.
A quantum phase space is distinguished by the canonical commutation relation between the quadratures, $[\hat{\vec{\Gamma}}_i, \hat{\vec{\Gamma}}_j] = \iu (\Omega_N)_{ij}$, with the matrix
\begin{equation}
\Omega_N = \begin{pmatrix}
	0 & \mathbb{I}_N \\
	-\mathbb{I}_N & 0
\end{pmatrix}
,\end{equation}
defining the symplectic structure of the phase space and $\mathbb{I}_N$ being the $N \times N$ unit matrix.
In this work we will represent the multimode quantum state in phase space using the Wigner function $W(\vec{\gamma})$.
Starting from the density operator $\hat{\rho}$ of a quantum state, we can introduce the characteristic function $\chi(\vec{\beta}) = \tr[\hat{\rho} \hat{D}(\vec{\beta})]$ as the expectation value of the multimode displacement operator $\hat{D}(\vec{\beta}) =  \exp(-\iu \hat{\vec{\Gamma}}^\T \Omega \vec{\beta})$ for the quantum state.
The multimode Wigner function is the (symplectic) Fourier transform of the characteristic function \cite{Adesso2014},
\begin{equation}
	W(\vec{\beta}) = \frac{1}{(2\pi)^{2N}}\int_{\mathbb{R}^{2N}} \exp(\iu \vec{\gamma}^\T \Omega_N \vec{\beta}) \chi(\vec{\beta}) \dd^{2N} \beta
.\end{equation}
The Wigner function can take on negative values and thus cannot be interpreted as a probability distribution.
The phase-space representation of a multimode quantum state proves to be very resourceful, when dealing with a unitary time evolution based on a Hamiltonian quadratic in the quadrature operators of $\hat{\vec{\Gamma}}$, as is the case for a $\chi^{(2)}$-nonlinear interaction.
Assuming a quadratic Hamiltonian and a closed system, the time evolved quadratures $\hat{\vec{\Gamma}}_{\text{out}} = M \hat{\vec{\Gamma}}_{\text{in}}$ are related to the input quadratures, $\hat{\vec{\Gamma}}_{\text{in}}$, by a symplectic matrix $M$.
The symplectic matrices are defined by the property $M^\T \Omega_N M = \Omega_N$.
Thus, the Wigner function of the output quantum state $W_{\text{out}}$, after time evolution under a quadratic Hamiltonian, is related to the input quantum state Wigner function, $W_{\text{in}}$, by $W_{\text{out}}(\vec{\gamma}) = W_{\text{in}}(M^{-1}\vec{\gamma})$, if the system is closed.
The symplectic matrices allow for a special singular value decomposition, $M = O Z O^\prime$, with $O$, $O^\prime$ being orthogonal symplectic matrices and $Z$ being a diagonal symplectic matrix which squeezes one quadrature by a factor and the corresponding orthogonal quadrature by the inverse factor.
This decomposition is called the Bloch-Messiah decomposition (BMD).
The matrices $O$, $O^\prime$ correspond to the passive, multiport beam splitting action, while $Z$ corresponds to the active single-mode squeezing action of the time evolution mediated by $M$ \cite{Houde2024a,Weedbrook2012,Adesso2014,Bloch1962,Braunstein2005a}.
In the next section we will show that a second-order nonlinear interaction of a pulsed open quantum system cannot always be described by a symplectic transformation and we thus need to generalize the formalism presented in this section.

\section{Operator input-output relation}\label{sec:in_out_op}
The general setting we consider consist of a pulsed quantum state of light as the input to an optical element with a second-order nonlinearity, which transforms the statistics and mode function of the input into an output quantum state residing in potentially different mode functions.
Our goal is to find the output quantum state in a predetermined set of broadband output modes, given an input quantum pulse and the input-output relation connecting the quadrature operators before and after the nonlinear interaction.
We denote the operators annihilating a photon of continuous (angular) frequency $\omega$ in the Fock space of the optical system at the input and output of the nonlinear optical element with $\hat{a}_{\text{in}}(\omega)$ and $\hat{b}_{\text{out}}(\omega)$ respectively.
Furthermore, we assume the input and output mode operators are related by the Bogoliubov transformation,
\begin{align} \label{eq:output_input_relation}
	\b_\text{out}(\omega) &= \int_0^\infty F(\omega, \Omega) \a_\text{in}(\Omega) \dd\Omega \nonumber \\
	&+ \int_0^\infty G^*(\omega, \Omega) \a_\text{in}^\dagger(\Omega) \dd\Omega,
\end{align}
linear in creation and annihilation operator \cite{Wasilewski2006,Christ2013,parametric-amplification-quantum-pulse}. The integration kernels $F$ and $G$ conserve the canonical commutation relation $\delta(\omega - \omega') = [\hat{b}_\text{out}(\omega), \hat{b}_\text{out}^\dagger(\omega')]$.
The Bogoliubov transformation in Eq.~\eqref{eq:output_input_relation} can result from a Hamiltonian quadratic in the creation and annihilation operators as shown in \cite{Wasilewski2006,Guedes2019,Christ2013} and Appendix~\ref{a:conversion}.
Defining quantum states in the Hilbert space of continuous frequency can be intricate, e.g. leading to divergences.
For this reason, it can be advantageous to discretize the input-output relation in Eq.~\eqref{eq:output_input_relation} using broadband modes (also known as temporal modes).
One such broadband mode basis can be generated using the Bloch-Messiah decomposition (BMD), reducing Eq.~\eqref{eq:output_input_relation} to the squeezing of individual broadband modes \cite{Houde2024a,Braunstein2005a,Wasilewski2006,Law2000}.
For some cases it might be sufficient to only consider a few of these most significant broadband eigenmodes of the BMD.
However, in general, the basis given by the BMD will not be the most efficient one, when describing the transformation of the quantum state in an arbitrary input mode $u(\omega)$, and we must resort to a description of only the relevant (and possibly different) input and output modes \cite{parametric-amplification-quantum-pulse}.
For example, the first Bloch-Messiah mode of the nonlinear interaction in a thin nonlinear crystal, as discussed in Sec.~\ref{sec:first_order_unitary}, is centered at half the central frequency of the classical pulse driving the interaction \cite{Hubenschmid2025}.
However, in electro-optic sampling one is interested in the up-conversion of low frequency pulses to frequencies overlapping with the classical pulse.
Thus, a BMD does not result in a resourceful description.

In the scenario we consider, the input quantum state occupies $N$ broadband modes with operators $\a_{u_i} = \int \dd \omega u_i^\ast(\omega)\a_\text{in}(\omega)$, while all other orthogonal modes are in the vacuum, and the $M$ relevant output modes are described by $\b_{v_i} = \int \dd \omega v_i^\ast(\omega)\b_\text{out}(\omega)$.
At this point the output modes are completely arbitrary. 
However, they could be determined by further processing, e.g., by the local oscillator pulse of a homodyne detection, or be obtained using the formalism presented in \cite{parametric-amplification-quantum-pulse}.
By inserting the input-output relation of Eq.~\eqref{eq:output_input_relation} into the definition of the output modes, we can relate
\begin{align}\label{eq:output_operator}
    \b_{v_i} &= \zeta_i\a_{f_i} + \xi_i\a_{g_i}^\dagger
\end{align}
to the (possibly overlapping) mode functions
\begin{align}
    f_i(\Omega) &= \frac{1}{\zeta_i}\int F^\ast(\omega, \Omega) v_i(\omega) \dd\omega, \\
    g_i(\Omega) &= \frac{1}{\xi_i}\int G^\ast(\omega, \Omega) v_i^\ast(\omega) \dd\omega,
\end{align}
normalized through $\zeta_i$ and $\xi_i$.
Since the modes $f_i$ and $g_i$ are possibly overlapping, they are not orthogonal with respect to the Hilbert-Schmidt norm and in turn, do not obey the canonical commutation relation.
Yet, to write the input-output relation in terms of the occupied input modes and some vacuum modes, we can utilize the Gram-Schmidt procedure to transform the linearly independent modes $\{u_1, \ldots, u_N, g_1, \ldots g_M, f_1, \ldots f_M\}$ into a set $\mathcal{B} = \{u_1, \ldots u_N, h_1, \ldots h_{2M}\}$ of orthogonal modes.
The linear independency is assumed here, but holds true for most practical examples.
In some cases it is possible to use a smaller set of modes than $\mathcal{B}$ \cite{parametric-amplification-quantum-pulse}, however here we focus on the most general case.
Inverting the Gram-Schmidt basis allows us to express the output modes, $v_i(\omega)$, with elements of the Gram-Schmidt basis $\mathcal{B}$.
To facilitate an input-output relation of the quantum state using the Wigner function, we introduce the quadrature operators $\hat{x}_{f} = \frac{1}{\sqrt{2}}(\hat{a}_f + \hat{a}_f^\dagger)$ and $\hat{p}_f = \frac{1}{\sqrt{2}\iu}(\hat{a}_f - \hat{a}_f^\dagger)$.
Organizing the quadratures of the output, input and orthogonal (vacuum) modes into the vectors
\begin{align}
	\hat{\Vec{\Gamma}}_\text{out} &= (\hat{x}_{v_1}, \ldots, \hat{x}_{v_M}, \hat{p}_{v_1}, \ldots, \hat{p}_{v_M})^\T, \label{eq:quad_out}\\
	\hat{\Vec{\Gamma}}_\text{in} &= (\hat{x}_{u_1}, \ldots, \hat{x}_{u_N}, \hat{p}_{u_1}, \ldots, \hat{p}_{u_N})^\T, \label{eq:quad_in}\\
	\hat{\Vec{\Gamma}}_\perp &= (\hat{x}_{h_1}, \ldots, \hat{x}_{h_{2M}}, \hat{p}_{h_1}, \ldots, \hat{p}_{h_{2M}})^\T\label{eq:quad_vac}
,\end{align}
allows recasting the input-output relation in terms of phase-space operators,
\begin{equation}\label{eq:in_out_quadratures}
	\hat{\Vec{\Gamma}}_\text{out} = A \hat{\Vec{\Gamma}}_\text{in} + B\vec{\Gamma}_{\perp}
.\end{equation}
The matrix
\begin{equation}
	A =  \begin{pmatrix}
		\Re(F_\text{in} + G_\text{in}) & \Im(-F_\text{in} + G_\text{in}) \\
		\Im(F_\text{in} + G_\text{in}) & \Re(F_\text{in} - G_\text{in})
	\end{pmatrix}
\end{equation}
is determined by the matrix elements $[F_\text{in}]_{ji} = \inprod{u_i}{f_j}^\ast = \iint v_j^\ast(\omega) F(\omega,\Omega)u_i(\Omega)\dd \omega \dd \Omega$ and $[G_\text{in}]_{ji} = \inprod{u_i}{g_j} = \iint v_j^\ast(\omega) G^\ast(\omega,\Omega) u_i^\ast(\Omega)\dd \omega \dd \Omega$, while $B$ depends on the matrix elements of $F$ and $G$ with respect to the modes $v_j$ and $h_i$ (see Appendix~\ref{a:in_out_relation} for details).
If no vacuum modes are involved, $B=0$, and the input-output relation of the quadrature operators conserves the cannonical commutation relation, the matrix $A$ in Eq.~\eqref{eq:in_out_quadratures} becomes symplectic.
In general, however, $A$ might not even by a square matrix.
To resolve this issue we introduce the generalized Bloch-Messiah decomposition in the following section.

\section{The generalized Bloch Messiah decomposition} \label{sec:gbmd}
If the number of input modes $N$ differs from the number of output modes $M$, the matrix $A$ in Eq.~\eqref{eq:in_out_quadratures} is not a square matrix (in general $A \in \mathbb{R}^{2M \times 2N}$).
Yet, the rectangular shape of $A$ suggest a more efficient mode basis for which a square matrix exists, only involving the $\min\{N, M\}$ input and $\min\{N, M\}$ output modes.
For this reason we introduce the \textit{generalized Bloch-Messiah decomposition} (GBMD).
As we show in Appendix~\ref{a:gbmd}, we can decompose
\begin{equation}\label{eq:gbmd}
	A = \begin{cases}
		S_L^\T P_L^\T \mathcal{A} & M > N \\
		\mathcal{A} & M = N \\
		\mathcal{A} P_R S_R & M < N \\
	\end{cases}
\end{equation}
with $S_L \in \mathbb{R}^{2M \times 2M}$, $S_R \in \mathbb{R}^{2N \times 2N}$ being symplectic, while $P_L \in \mathbb{R}^{2N \times 2M}$, $P_R \in \mathbb{R}^{2M \times 2N}$ reduce the dimensionality and $\mathcal{A}$ being a real square matrix with dimension $2\min\{M, N\}$.
The indices $L$ indicate the left-sided decomposition and $R$ the right-sided one.

We call Eq.~\eqref{eq:gbmd} the generalized Bloch-Messiah decomposition, since it is based on the symplectic version of the singular value decomposition of a rectangular matrix $A$ introduced by Xu \cite{Xu2003,Xu2005}.
Different to the usual Bloch-Messiah decomposition, $A$ does not need to be symplectic.
However, we assume $A \Omega A^\T$ is nonsingular.
This is usually the case as long as all output modes overlap with at least one input mode after it is transformed by the nonlinear interaction.
There exist an algorithm to compute the GBMD derived by Xu \cite{Xu2005} based on symplectic Housholder and Givens rotation.
We present an example of the algorithm in Appendix~\ref{a:gbmd_example} and the corresponding \texttt{Python} implementation can be found in \cite{SubcycleQ}.
The decomposition in Eq.~\eqref{eq:gbmd} has great potential for many settings in quantum optics.
The GBMD could be used to describe the single mode measurement of a multimode quantum state using electro-optic sampling (nonlinear homodyne detection) \cite{Hubenschmid2024,Kalash2023,Sendonaris2024} or to orthogonalize overlapping mode functions as proposed for the correlation tomography in \cite{Hubenschmid2025}.
The simplified input-output relation resulting from the GBMD allows us to formulate an input-output relation of the quantum states.
However, since the GBMD reduces the dimension of either the input or output phase space, the quantum-state input-output relation, derived in the following section, necessarily needs to be an open-systems description.

\section{The input-output relation in phase space}\label{sec:in_out_ps}
In this section we derive the phase-space input-output relation based on the multimode Wigner function of the input and output state.
The generalized Bloch-Messiah decomposition in Eq.~\eqref{eq:gbmd} allows us to find a set of $\min\{N,M\}$ effective modes, describing the transformation of the quantum state due to the nonlinear interaction.
The output Wigner function presents the main result of this work.
Since only one of the three cases $M > N$, $M = N$ or $M < N$ can be true at the same time, we will derive the phase-space input-output relation for all three cases separately.

\subsection{Input mode surplus ($M < N$)}
\begin{figure*}
	\centering
	\includegraphics[width=\textwidth]{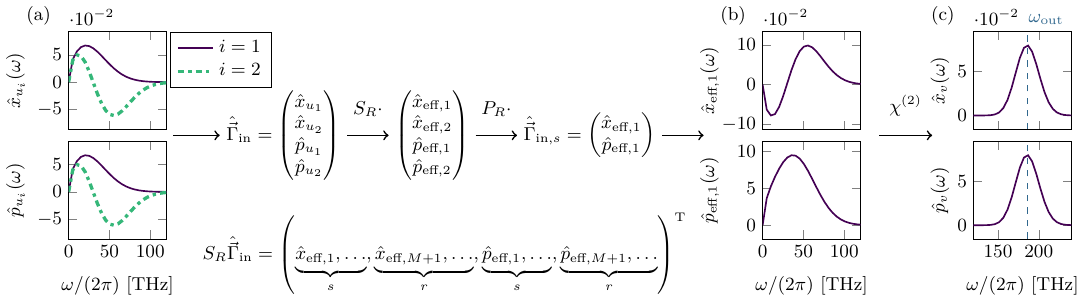}
	\caption{(a) The input modes of $\hat{\vec{\Gamma}}_\text{in}$ in the frequency domain.
        (b) The effective input modes, which are obtained from the generalized Bloch-Messiah decomposition in Eq.~\eqref{eq:gbmd}, by first transforming the input modes using the symplectic matrix $S_R$ and afterwards reducing the number of effective modes using $P_R$ to match the number, $M$, of output modes $v_i$ (in this case we consider one output mode, $M=1$, centered at $\omega_\text{out}$).
        We call the modes which are kept system modes (labeled by $s$) and label the residual modes by $r$.
        The effective modes are up-converted to the output modes in $\hat{\vec{\Gamma}}_\text{out}$, shown in (c), by the $\chi^{(2)}$-nonlinear interaction resulting in Eq.~\eqref{eq:in_out_quadratures}.
        Here we used the nonlinear interaction introduced in Sec.~\ref{sec:first_order_unitary} as an example.
	}
	\label{fig:modeTransformation}
\end{figure*}

In the case of an input-mode surplus ($M < N$), the generalized Bloch-Messiah decomposition is $A = \mathcal{A} P_R S_R$.
Thus, we can collect the effective system modes, i.e., the modes transformed into the relevant output modes by the nonlinear interaction, in $\hat{\Vec{\Gamma}}_{\text{in,s}} = P_R S_R \hat{\Vec{\Gamma}}_{\text{in}}$.
While $S_R$ transforms the phase space of the input modes, the matrix $P_R$ reduces the $2N$ components of the vector $S_R \hat{\Vec{\Gamma}}_{\text{in}}$ to the $2M$ components of the output phase space, as visualized in Fig.~\ref{fig:modeTransformation}.
We define the reduced input state of the system modes, $\hat{\rho}_{\text{in,s}} = \tr_r(\hat{\rho}_\text{in})$, by tracing over the remaining $N-M$ modes not contained in $\hat{\Vec{\Gamma}}_{\text{in,s}}$.
The modes traced out are marked by $r$ in Fig.~\ref{fig:modeTransformation} and will be called residual modes, as opposed to the remaining system modes marked by $s$.
Since only the system modes are relevant to the output quantum state, we introduce the Wigner function, $W_{\text{in}, s}(\vec{\gamma})$, of the reduced input quantum state, $\hat{\rho}_{\text{in,s}}$, of the system modes.
Furthermore, by defining the covariance matrix of the orthogonal (initial vacuum) modes, $\cov_{B} = BB^\T$, we can express the output Wigner function as a convolution of the multivariate Gaussian phase-space function,
\begin{align}
	G(\vec{\gamma}) &= \frac{2^M(2\pi)^{M}}{\sqrt{\det(\cov_{B})}} \exp(-\vec{\gamma}^\T\mathcal{A} \cov_B^{-1} \mathcal{A}^\T\vec{\gamma})
,\end{align}
and the input Wigner function, resulting in
\begin{equation}\label{eq:outputWignerfunction1}
	W_\text{out}(\vec{\gamma}) = \Big(W_{\text{in,s}} \ast G\Big)(\mathcal{A}^{-1}\vec{\gamma})
,\end{equation}
with $*$ denoting the convolution.
Thus, the reduced input Wigner function is smoothed out by the convolution with $G$ and transformed by $\mathcal{A}$.
A detailed derivation of Eq.~\eqref{eq:outputWignerfunction1} can be found in Appendix~\ref{a:inOutPhaseSpace}.
As a remark, it should be mentioned that for multiple nonlinear interactions in a setup, all Gaussian phase-space functions involved in the convolution can be collected into one Gaussian.
An example of Eq.~\eqref{eq:outputWignerfunction1} for a two-mode squeezed input pulse and a single (broadband) output mode is discussed in Sec.~\ref{ssec:squeezed}.

To better understand Eq.~\eqref{eq:outputWignerfunction1}, it is instructive to elaborate on a special case of the phase-space input-output relation.
Let us consider an input quantum pulse described by a Gaussian Wigner function,
\begin{equation}
    W_\text{in}(\vec{\gamma}) = \frac{1}{\pi^M\sqrt{\det(\cov_\text{in})}}\exp(-\vec{\gamma}^\T \cov_\text{in}^{-1}\vec{\gamma})
,\end{equation}
with covariance matrix $\cov_\text{in}$ and no displacement in phase space.
Since the nonlinear interaction is described by a quadratic Hamiltonian, the output state is still Gaussian with a covariance matrix
\begin{equation}\label{eq:gaussian_output}
	\cov_\text{out} = \cov_{B} + \mathcal{A}\cov_{\text{in,s}}\mathcal{A}^\T
,\end{equation}
and the covariance matrix $\cov_{\text{in,s}} = P_R S_R \cov_\text{in} S_R^\T P_R^\T$ corresponding to the covariance matrix of the reduced input state of the system modes.

\subsection{Output mode surplus ($M > N$)}
In the case of an output mode surplus ($M>N$), only a fraction of the output modes are necessary to represent the transformed input quantum state.
Utilizing the GBMD $A = S_L^\T P_L^\T \mathcal{A}$, we can define the phase space of the reduced output modes, i.e., the system modes, as $\vec{\gamma}_s = P_L S_L^{-1} \vec{\gamma}$ and the phase space of the environment $\vec{\gamma}_e$, which together comprise the phase space of the full output Wigner function, $\vec{\gamma}$.
Furthermore, by separating the covariance matrix of the orthogonal modes into system, $\cov_{B,ss} = P_L S_L \cov_B S_L^\T P_L^\T$, and an environment part, $\cov_{B; ee}$, with the latter corresponding to the remaining blocks on the diagonal of $S_L \cov_B S_L^\T$ besides $\cov_{B, ss}$ (see Appendix~\ref{a:inOutPhaseSpace} for more details).
If we transform to the basis $T \hat{\vec{\Gamma}}_\perp = (\hat{x}_{h_1}, \hat{p}_{h_1}, \hat{x}_{h_2}, \hat{p}_{h_2}, \ldots, \hat{x}_{h_{2M}}, \hat{p}_{h_{2M}})^\T$, the transformed covariance matrix disintegrates into the four blocks,
\begin{equation}
    T S_L \cov_B S_L^\T T = \begin{pmatrix}
        \cov_{B,ss} & \cov_{B,se} \\
        \cov_{B,es} & \cov_{B,ee} \\
    \end{pmatrix}
,\end{equation}
with the system and environment blogs on the diagonal.
By introducing the  Schur complement, $\cov_{\text{schur}} = \cov_{B,\text{ss}} - \cov_{B,\text{se}} \cov_{B,\text{ee}}^{-1}\cov_{B,\text{es}}$, together with $\vec{\gamma}_\text{schur} = \cov_{B,\text{es}}\cov_{B,\text{ee}}^{-1}\vec{\gamma}_\text{e}$, connecting the system and environment by the off-diagonal blocks $\cov_{B,\text{se}} = \cov_{B,\text{es}}^\T$, we can again express the output Wigner function by a convolution.
Compared to the case in the previous section, the multivariate Gaussian phase-space function,
\begin{align}
	G(\vec{\gamma}_\text{s}) &= \frac{2^M}{(2\pi)^{M - 2L}}\left[\det(\cov_{\text{schur}})\det(\cov_{B,\text{ee}})\right]^{-\frac{1}{2}} \nonumber \\
	&\qquad\times \exp(-\vec{\gamma}_\text{s}^\T\mathcal{A} \cov_{\text{schur}}^{-1} \mathcal{A}^\T\vec{\gamma}_\text{s}),
\end{align}
is determined by Schur complement $\cov_\text{Schur}$ of the covariance matrix $\cov_B$ with respect to the system modes.
The output Wigner function,
\begin{equation}
	W_\text{out}(\vec{\gamma}) = \Big(W_{\text{in}} \ast G\Big)(\mathcal{A}^{-1}[\vec{\gamma}_\text{s} + \vec{\gamma}_\text{schur}])
,\end{equation}
is again the input Wigner function, but smoothed out by the convolution with $G$ and rescaled by $\mathcal{A}$, however, additionally displaced by $\vec{\gamma}_\text{schur}$.

\subsection{Equal mode number ($M = N$)}\label{ssec:equal_mode_num}
If the number of input modes equals the number of output modes, we cannot reduce the dimension of the input state nor the output state.
However, this is not necessary since the matrix $A$, describing the nonlinear interaction between in- and output modes, is a square matrix.
Thus, the GBMD is not necessary.
The transformation $A$ given by the nonlinear process is still not symplectic, as long as the vacuum modes $h_i(\omega)$ are involved.
Yet, we can describe the phase-space input-output relation using the Gaussian phase-space function
\begin{align}\label{eq:gaussian_equal_mode_num}
	G(\vec{\gamma}) &= \frac{2^{2M}\pi^{M}}{\sqrt{\det(\cov_{B})}} \exp(-\vec{\gamma}^\T A \cov_B^{-1} A^\T\vec{\gamma})
\end{align}
by
\begin{equation}
	W_\text{out}(\vec{\gamma}) = \Big(W_{\text{in}} \ast G\Big)(A^{-1}\vec{\gamma})
.\end{equation}
This case will be further discussed in Sec.~\ref{ssec:fock} for a pulse of a single (broadband) mode Fock state ($N=M=1$).

\section{THz to optical-frequency conversion}\label{sec:first_order_unitary}
\begin{figure}
	\centering
	\includegraphics[width=0.45\textwidth]{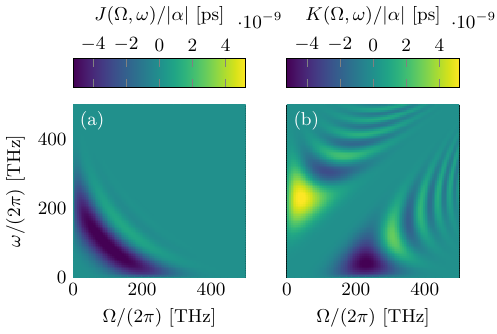}
	\caption{
		An example of the kernels $J(\Omega, \omega)$ and $K(\Omega, \omega)$ describing the nonlinear interaction mediating between the input and output modes, which is defined in Eq.~\eqref{eq:nl_unitary_first_order}.
		The kernel $J(\Omega, \omega)$ describes the strength of a squeezing interaction between the continuous (angular) frequency mode at $\Omega$ and the mode at $\omega$, simultaneously creating or annihilating a photon in each mode.
		The kernel $K(\Omega, \omega)$ corresponds to a beam splitter interaction between the continuous frequency modes and creates a photon in the mode $\omega$, while annihilating one at $\Omega$.
		The nonlinear crystal is assumed to be zinc tellurid with a refractive index $n(\omega)$ \cite{Marple1964}, length $L = \SI{20}{\micro\meter}$ and an electro-optic coefficient $r_{41} = \SI{4}{\cdot 10^{-12}\second\ampere\per\kilo\gram\meter}$ \cite{Boyd2019}.
		The classical coherent pulse driving the nonlinear interaction is assumed to be centered at $\omega_\text{p} / (2\pi) = \SI{200}{\tera\hertz}$, to have a bandwidth of $\Delta \omega_{\text{p}} / (2\pi) = \SI{118}{\tera\hertz}$ and a coherent amplitude of $\alpha$.
		We assume the nonlinear crystal is in free-space, the coherent pulse is in the fundamental paraxial mode with beam waist area $A=\SI{28}{\micro\meter\squared}$ and polarized orthogonal to the mode of the quantum interaction.
		Note that for one frequency, $\Omega$ or $\omega$, fixed to $\omega_\text{p}$, most of the contribution from the squeezing interaction is below the central frequency of the coherent pulse, but above for the beam splitting contribution.
	}
	\label{fig:nl_interaction_kernel}
\end{figure}

To illustrate the input-output theory derived above, we consider the frequency conversion of THz radiation to optical frequencies using a zinc-tellurid nonlinear crystal in free space, as often encountered in electro-optic sampling \cite{Namba1961,Leitenstorfer1999}.
With a crystal orientation as in \cite{Moskalenko2015}, the effect of the second-order nonlinear susceptibility $\chi^{(2)}$ is described by a unitary time-evolution operator,
\begin{align}\label{eq:nl_unitary_first_order}
\begin{split}
	\hat{U}_{\text{NL}} &= \exp\Bigg(\frac{1}{2}\iint_0^\infty J(\Omega, \omega) \hat{a}_\omega \hat{a}_\Omega \dd \omega \dd \Omega - \hc \\
	&\qquad + \iint_0^\infty K(\Omega, \omega) \hat{a}_\omega^\dagger \hat{a}_\Omega \dd \omega \dd \Omega \Bigg),
\end{split}
\end{align}
quadratic in the operators acting on the modes polarized orthogonal to the strong classical pulse driving the nonlinear interaction.
The kernels $J(\Omega, \omega)$ and $K(\Omega, \omega)$, accounting for a squeezing and beam-splitting interaction respectively, are given in Eq.~\eqref{eq:squeezing_nl} and \eqref{eq:beam_splitting_nl}.
The squeezing interaction simultaneously creates/annihilates a photon at frequency $\omega$ and $\Omega$, while the beam splitting interaction creates a photon at frequency $\omega$ and annihilating a photon at frequency $\Omega$.
Both kernels scale linearly with the coherent amplitude $\alpha$ of the classical pulse driving the nonlinear interaction.
An example of the kernels for a thin crystal of length $L = \SI{20}{\micro\meter}$ and a probe pulse centered at $\SI{200}{\tera\hertz}$ with a bandwidth of $\SI{118}{\tera\hertz}$ are shown in Fig.~\ref{fig:nl_interaction_kernel}.

Using the first-order unitary introduced in \cite{Onoe2022, Hubenschmid2024} and defining $f_K(\Omega) = -\int K^\ast(\Omega, \omega) v(\omega) \dd \omega$, $f_J(\Omega) = \int J^\ast(\Omega, \omega) v^\ast(\omega) \dd \omega$ we can calculate the parameters
\begin{align}
	\theta_K &= \left(\iint_0^\infty \abs{f_K(\Omega)}^2 \dd \Omega\right)^{1/2}, \\
	\theta_J &= \left(\iint_0^\infty \abs{f_J(\Omega)}^2 \dd \Omega\right)^{1/2}
,\end{align}
determining the amount of beam splitting and squeezing involved in the nonlinear interaction which results in a given output mode $v(\omega)$.
Depending on whether beam splitting or squeezing dominates, the parameter $\theta = \sqrt{|\theta_K^2 - \theta_J^2|}$ defines a beam-splitting or squeezing relation,
\begin{equation}\label{eq:mu_nu}
	(\mu, \nu) = \begin{cases}
		(\cos[\theta], \sin[\theta]) & \theta_K > \theta_J \\
		(\cosh[\theta], \sinh[\theta]) & \theta_K < \theta_J
	\end{cases}
,\end{equation}
resulting in $F(\omega, \Omega) = \mu \delta(\omega - \Omega) - \nu K(\Omega, \omega) / \theta$ and $G(\omega, \Omega) = \nu J(\Omega, \omega) / \theta$.
The first-order unitary can be understood as a first-order perturbation theory (using the Baker-Campbell-Hausdorff formula) renormalized to fulfill the bosonic commutation relation.
We will refer to the parameter range with $\theta_K > \theta_J$ as the beam-splitting regime, while $\theta_K < \theta_J$ is called the squeezing regime.
A simplified version of single (monochromatic) mode frequency conversion in the beam splitter regime can be found in \cite{Tucker1969} and for a broadband but mode selective case in \cite{Quesada2016}.
The mode function $v(\omega)$ could be determined by using the formalism introduced by Tziperman \emph{et al.} \cite{parametric-amplification-quantum-pulse}.
However, here we follow an example suited for electro-optic sampling.
In this case, the local oscillator pulse in the optical frequencies used in the homodyne detection subsequent to the nonlinear interaction determines the shape of $v(\omega)$, while the input quantum pulse is in the THz range and is up-converted by the nonlinear interaction.
A similar scenario occurs in nonlinear homodyne detection \cite{Kalash2023,Kalash2025}, mode selective detection \cite{Ansari2018,Kouadou2023} or nonlinear photo-detection \cite{Yanagimoto2023a,Sendonaris2024}.
In all these examples, the nonlinear interaction between the input and output quantum pulse is (in the most general case) not described by a symplectic transformation.

\section{Examples} \label{sec:examples}
In these sections we will investigate two examples of the phase-space input-output relation.
First we consider a Fock state of a single broadband mode in the THz frequencies up-converted to a single mode in the optical frequencies, where it might be detected or further processed.
Second, we investigate the up-conversion of broadband two-mode squeezed state in the THz range to one optical broadband mode.
In both cases we assume the central frequency of the output mode $\omega_\text{out}$, as well as the strength of the nonlinear interaction given by the amplitude $\alpha$ of the driving pulse to be variable.

\subsection{Single mode Fock state}\label{ssec:fock}
The scenario we consider in this section consist of one broadband mode $u(\omega)$ in the THz regime [c.f. $u_1(\omega)$ in Fig.~\ref{fig:modeTransformation} (a)], which is occupied by the $n$-photon Fock state, while the output mode $v(\omega)$ is in the optical frequency range [c.f. Fig.~\ref{fig:modeTransformation} (c)].
The broadband mode of the Fock state is centered at $\omega_\text{in} = \SI{27}{\tera\hertz}$ and has a frequency spread of $\SI{35}{\tera\hertz}$, while the output mode $v(\omega)$ has a fixed bandwidth $\Delta \omega / (2\pi) = \SI{24}{\tera\hertz}$ and a central frequency either at $\omega_{\text{out}} / (2\pi) = \SI{184}{\tera\hertz}$ or $\SI{200}{\tera\hertz}$.

\begin{figure}
	\centering
	\includegraphics[width=0.45\textwidth]{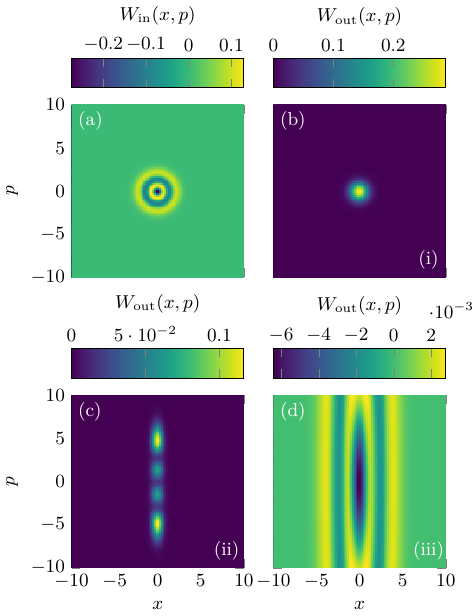}
	\caption{
		(a) Wigner function of a pulsed THz three-photon Fock state used as the input to the nonlinear interaction.
		(b) and (c) show the output Wigner function for the parameter sets (i) and (ii) defined in the main text, and visualized in Fig \ref{fig:oneInOneOut}, corresponding to the beam-splitter regime introduced in Sec.~\ref{sec:first_order_unitary}.
		In (b) the amplitude $\alpha$ of the coherent pulse driving the nonlinear interaction is chosen such that no up-conversion of the THz quantum pulse to the optical output mode occurs, showing the non-monotonic dependence of the up-conversion on the amplitude $\alpha$ in the beam-splitter regime.
		Thus the Wigner function corresponds to the vacuum of the optical modes.
		While in (c) the point of optimal up-conversion for a given $\omega_\text{out}$ is used, resulting in an output Wigner function resembling the quadrature distribution of a Fock state, i.e., a Hermite-Gauss distribution.
		The parameters of (d) are in the squeezing regime, where no optimal point of up-conversion exists.
		The stronger the amplitude of the coherent pulse, $\alpha$, the more the Fock state gets amplified in phase space.
	}
	\label{fig:fock_out}
\end{figure}

Since we only have one input and one output mode, we can use the expression in Sec.~\ref{ssec:equal_mode_num}.
Using a singular value decomposition, the matrix $A^{-1} \cov_B [A^{-1}]^\T = \sigma_x \vec{e}_x \vec{e}_x^\T + \sigma_p \vec{e}_p \vec{e}_p^\T$ is diagonal in the phase space basis $\vec{e}_x$, $\vec{e}_p$ with singular values $\sigma_x$, $\sigma_p$.
By defining $A^{-1}\vec{\gamma} = y\vec{e}_x + z\vec{e}_p$ we can express the Wigner function of the output state for an $n$-photon input state as,
\begin{widetext}
\begin{equation}\label{eq:fock_out}
	W_\text{out}(\vec{\gamma}) = \frac{\exp(-\frac{z^2}{\sigma_x + 1}-\frac{y^2}{\sigma_p + 1})}{\pi\abs{\det(A)}\sqrt{(\sigma_x + 1)(\sigma_p + 1)}} \sum_{m=1}^{n} \left(\frac{\sigma_x - 1}{\sigma_x + 1}\right)^{m}L^{(-\frac{1}{2})}_{m}\left(\frac{-2 z^2}{\sigma_x^2-1}\right) \left(\frac{\sigma_p - 1}{\sigma_p + 1}\right)^{n-m}L^{(-\frac{1}{2})}_{n-m}\left(\frac{-2 y^2}{\sigma_p^2-1}\right)
.\end{equation}
\end{widetext}
Examples of Eq.~\eqref{eq:fock_out} for a three photon input state can be seen in Fig.~\ref{fig:fock_out} using the nonlinear process presented in Sec.~\ref{sec:first_order_unitary}.
The three-photon input state ($n = 3$) is shown in Fig.~\ref{fig:fock_out} (a), while the output state calculated according to Eq.~\eqref{eq:fock_out} is visualized in Fig.~\ref{fig:fock_out} (b)-(d) for different combinations of the output modes central frequency $\omega_\text{out}$ and coherent amplitude $\alpha$ determining the strength of the nonlinear interaction.

\begin{figure}
	\centering
	\includegraphics[width=0.45\textwidth]{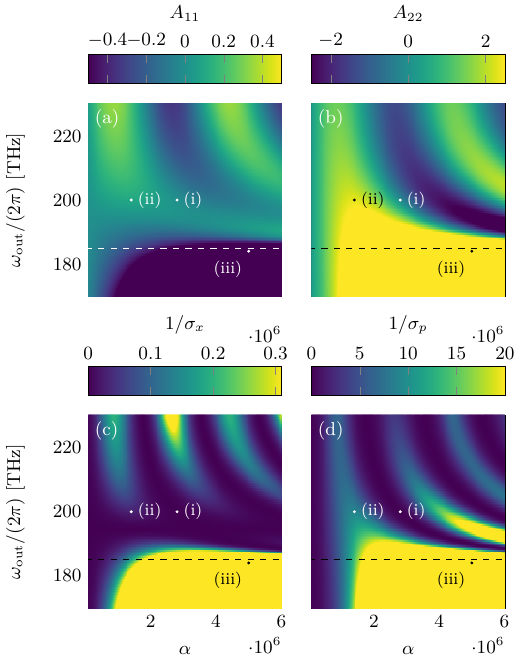}
	\caption{
		The four parameters, independent of the input quantum state, determining the Wigner function of the output quantum state, as a function of the central frequency of the output mode $\omega_\text{out}$ and the amplitude $\alpha$ of the coherent pulse driving the nonlinear interaction.
		(a), (b) The nonzero elements $A_{11}$, $A_{22}$ of the matrix $A$, used in the input-output relation of Eq.~\eqref{eq:in_out_quadratures}.
		The matrix $A$ scales the phase space of the output Wigner function.
		(c), (d) The singular values $\sigma_x$ and $\sigma_p$ of the covariance matrix defining the Gaussian phase-space function in Eq.~\eqref{eq:gaussian_equal_mode_num} as a function of the same parameters.
		Since $\sigma_x$, $\sigma_p$ determine the width of the Gaussian, the input Wigner function is convolved with, they determine the smoothness of the output Wigner function.
		The inverse of $\sigma_x$ and $\sigma_p$ are plotted to avoid divergences.
		The horizontally dashed lines mark the transition from the squeezing to the beam-splitting regime, introduced in Sec.~\ref{sec:first_order_unitary}, as $\omega_\text{out}$ is increased.
        The colormap is saturated in the squeezing regime. 
	}
	\label{fig:oneInOneOut}
\end{figure}

It is important to note here, that in general the time evolution connecting the input and output Wigner function is not unitary and the system of input and output modes is not closed since the vacuum modes in Eq.~\eqref{eq:in_out_quadratures} are traced out during the calculation of the output quantum state.
Tracing over the additional modes will admix some of the (transformed) vacuum modes to the output quantum state which is quantified by the singular values $\sigma_x$, $\sigma_p$.
Due to the additional contributions from the traced out vacuum modes, $A$ does not need to be symplectic to ensure the conservation of the canonical commutation relation.
As is evident from Eq.~\eqref{eq:fock_out}, the output Wigner function is determined by the matrix $A$ and the singular values $\sigma_x$ and $\sigma_p$, which in turn depend on the central frequency of the output mode, $\omega_\text{out}$, and the coherent amplitude, $\alpha$,  of the pulse driving the interaction.
In the case of $\alpha \in \mathbb{R}$ and $v(\omega), u(\omega) \in \mathbb{R}$ for all $\omega$, the matrix $A$ becomes diagonal and we can represent the dependence of the two matrix elements $A_{11}$, $A_{22}$ and singular values $\sigma_x$, $\sigma_p$ on $\omega_\text{out}$, $\alpha$ in Fig.~\ref{fig:oneInOneOut}.

A prominent feature of Fig.~\ref{fig:oneInOneOut} is the different dependence on $\alpha$ in the beam-splitting and squeezing regime, introduced in Sec.~\ref{sec:first_order_unitary}.
While the central frequency of the output mode determines the regime, with a transition from squeezing to beam-splitting at $\omega_\text{out} / (2\pi) = \SI{185}{\tera\hertz}$, the parameter $\alpha$ affects the squeezing and beam splitting interaction the same and thus is proportional to $\theta$.
Therefore, the quantities determining the input-output relation of the quadratures, shown in Fig.~\ref{fig:oneInOneOut}, exhibits a periodic dependence on the coherent amplitude $\alpha$ in the beam splitting regime and exponential dependence on $\alpha$ in the squeezing regime.
It is an open question, if the transition from beam-splitting to squeezing regime is a phase transition, which will not be addressed in this work.

Having characterized the two different regimes of the up-conversion, we can explain the various aspects of the output Wigner function in Fig.~\ref{fig:fock_out}.
For the parameters $\omega_\text{out} / (2\pi) = \SI{200}{\tera\hertz}$ and $\alpha = 2.8 \cdot 10^{6}$ marked by (i) in Fig.~\ref{fig:oneInOneOut}, there is no up-conversion, since for these parameters $\theta = \pi$ and according to Eq.~\eqref{eq:first_order_in_out} the nonlinear interaction is described by the kernels $F(\omega, \Omega) = \delta(\omega - \Omega)$ and $G(\omega, \Omega) = 0$, restricting the interaction to equal frequencies.
This linear response is reflected in singular scaling parameters of the Wigner function $[A^{-1}]_{11}$ and $[A^{-1}]_{22}$, which remove the dependence of the output Wigner function on the input state.
The contribution visible in Fig.~\ref{fig:fock_out} (b) is purely due to the contribution of the vacuum modes with covariance matrix $\cov_{B} \approx \mathbb{I}$.

The parameters $\omega_\text{out} / (2\pi) = \SI{200}{\tera\hertz}$ and $\alpha = 1.4 \cdot 10^{6}$ marked by (ii) in Fig.~\ref{fig:oneInOneOut} are still in the beam-splitting regime, however, resulting in $\theta = \pi / 2$ and thus leading to the most efficient up-conversion of the THz excitations to the optical frequencies.
However, since the mode which is up converted to $v(\omega)$ does not match the input mode $u(\Omega)$ perfectly, there is some smoothing in the $p$ direction of the phase space, $\sigma_p = 0.11$, while substantial smoothing in the $x$ direction, $\sigma_x = 37.97$, suppresses the negativity in the output Wigner function, as can be seen in Fig.~\ref{fig:fock_out} (c).
Additionally the output Wigner function is deamplified in the $x$ direction by a factor $[A^{-1}]_{11} = 10.60$ and amplified by $[A^{-1}]_{22} = 0.43$ in the $p$ direction.

In the case of $\omega_\text{out} / (2\pi) = \SI{184}{\tera\hertz}$ and $\alpha = 5 \cdot 10^{6}$ marked by (iii) in Fig.~\ref{fig:oneInOneOut}, the nonlinear interaction is in the squeezing regime, amplifying both quadratures, i.e., $[A^{-1}]_{11} = -0.59$ and $[A^{-1}]_{22} = 0.05$.
Significantly, the smoothing in both directions is greatly reduced compared to case (ii), since $\sigma_x = 0.114$ and $\sigma_p = 0.002$, resulting in some negativity of the output Wigner function, visible in Fig.~\ref{fig:fock_out} (d).
The resilience of the Wigner negativity in the squeezing regime could originate from the broad frequency spectrum which is correlated to the output mode centered at $\omega_\text{out} / (2\pi) = \SI{184}{\tera\hertz}$ due to the squeezing interaction, $J(\Omega, \omega)$, in comparison to the beam splitting interaction, $K(\Omega, \omega)$, as seen in Fig.~\ref{fig:nl_interaction_kernel}.

Our results open a path to a possible optimization in the frequency conversion of pulsed quantum states exhibiting Wigner negativity.
By combining the input quantum state in the mode $u(\Omega)$ with a squeezed vacuum pulse in the mode $h_1(\Omega)$ orthogonal to $u(\Omega)$, the smoothing parameter in one direction, e.g., $\sigma_p$, could be reduced by the expense of enlarging the other.
The combined effect still could increase the total Wigner negativity in the output state.
We elaborate on the extension to multiple input modes using a two-mode squeezed state in the following section.

\subsection{Two-mode squeezed vacuum} \label{ssec:squeezed}
In the example considered in this section, two broadband modes in the THz regime, containing a two-mode squeezed vacuum state, are up-converted to the optical frequency range, in which only one mode is of further interest, as depicted in Fig.~\ref{fig:modeTransformation}.
Multimode squeezed states posses a Gaussian Wigner function, thus we can utilize Eq.~\eqref{eq:gaussian_output} to describe the output Wigner function.

\begin{figure}
	\centering
	\includegraphics[width=0.45\textwidth]{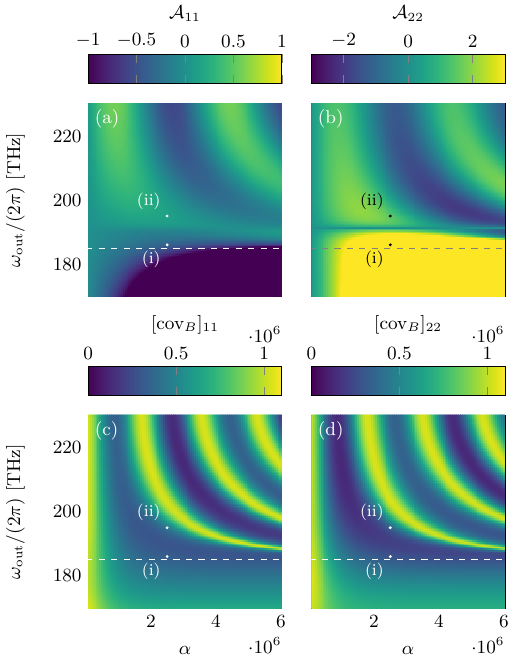}
	\caption{
		The four parameters, independent of the input quantum state, determining the Wigner function of the output quantum state as a function of the central frequency $\omega_\text{out}$ of the output mode and the amplitude $\alpha$ of the classical pulse driving the nonlinear interaction.
		(a), (b) The nonzero elements $\mathcal{A}_{11}$, $\mathcal{A}_{22}$ of the matrix $\mathcal{A}$. The matrix $\mathcal{A}$ is calculated using the generalized Bloch-Messiah decomposition as described in Sec.~\ref{ssec:squeezed}.
		The matrix $\mathcal{A}$ scales the phase space of the output Wigner functions.
		(c), (d) The two nonzero entries of the of the covariance matrix $\cov_B$ in Eq.~\eqref{eq:gaussian_output}, independent of the input states covariance matrix, as a function of the same parameters.
		The horizontally dashed lines mark the transition from the squeezing to the beam-splitting regime, introduced in Sec.~\ref{sec:first_order_unitary}, as $\omega_\text{out}$ is increased.}
	\label{fig:twoInOneOut}
\end{figure}

Since the number of output modes is smaller than the number of input modes, $M < N$ we need to find the GBMD, introduced in Sec.~\ref{sec:gbmd}, $A = \mathcal{A} P_R S_R$.
A \texttt{Python} implementation of the algorithm, presented in \cite{Xu2005} and described in Sec.~\ref{sm:example_gbmd}, for the specific setting considered in this section, can be found in \cite{SubcycleQ}. 
We use the same parameters for the non-linear crystal as in the previous example (and introduced in sec. \ref{sec:first_order_unitary}). 
As before, we assume the amplitude $\alpha$ of the classical pulse is real, resulting in a diagonal effective transformation matrix $\mathcal{A}$.
The diagonal entries of $\mathcal{A}$ are shown in Fig.~\ref{fig:twoInOneOut} (a) and (b) as a function of the central frequency of the output mode $\omega_{\text{out}}$ and the coherent amplitude driving the nonlinear interaction.
The matrix $\cov_B$ is diagonal as well, and the diagonal elements are shown in Fig.~\ref{fig:twoInOneOut} (c) and (d) as a function of the same parameters, $\omega_\text{out}$ and $\alpha$.
The two parameter sets exemplified in this section are marked by (i) for $\omega_\text{out} = \SI{186}{\tera\hertz}$ and $\alpha = 2.5 \cdot 10^6$ or (ii) for $\omega_\text{out} = \SI{195}{\tera\hertz}$ and $\alpha = 2.5 \cdot 10^6$.

\begin{figure}
	\centering
	\includegraphics[width=0.45\textwidth]{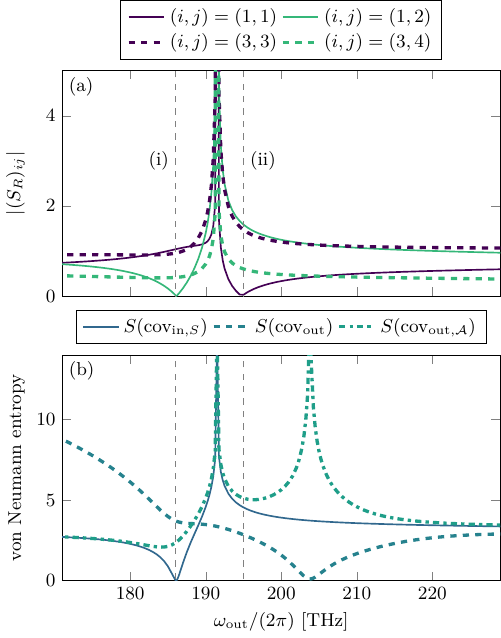}
	\caption{
		(a) The nonzero elements of the symplectic matrix $S_R$ transforming the quadratures of the input quantum state as a function of the central frequency $\omega_\text{out}$ of the coherent pulse driving the nonlinear interaction.
		The entries $(S_R)_{11}$ and $(S_R)_{12}$ determine the amount $\hat{x}_{u_1}$ and $\hat{x}_{u_2}$ contribute to the effective input quadrature $\hat{x}_\text{eff}$ shown in Fig.~\ref{fig:modeTransformation}, while $(S_R)_{33}$ and $(S_R)_{34}$ determine the amount $\hat{p}_{u_1}$ and $\hat{p}_{u_2}$ contribute $\hat{p}_\text{eff}$.
		(b) The von Neumann entropy of Gaussian states as a function of the central frequency $\omega_\text{out}$ of the output mode, calculated according to \cite{Holevo1999,Adesso2014,Weedbrook2012}.
		The von Neumann entropy is shown for three different cases: the reduced input state after transformation with $S_R$ (solid), the output state (dashed), and the output state after transforming the Wigner function with $\mathcal{A}$ (dash-dotted).
	}
	\label{fig:overlap}
\end{figure}

The most significant difference to the previous example in Sec.~\ref{ssec:fock}, is the reduction of the input modes to one effective input mode with quadrature operators $\hat{\vec{\Gamma}}_{\text{in,s}} = P_R S_R \hat{\vec{\Gamma}}_\text{in} = (\hat{x}_\text{eff}, \hat{p}_\text{eff})^\T$.
The effective input mode corresponding to an output mode with central frequency $\omega_\text{out} = \SI{195}{\tera\hertz}$ is shown in Fig.~\ref{fig:modeTransformation} (b).
While the $p$-quadrature is mostly independent of $\omega_\text{out}$, the $x$-quadrature changes drastically.
This is visible in Fig.~\ref{fig:overlap} (a), which shows how much the modes $u_1(\Omega)$ and $u_2(\Omega)$ contribute to the effective input quadratures as a function of the output modes central frequency.
The contribution of each input mode is determined by $S_R$ since the effective input mode corresponds to the first mode in $S_R \hat{\vec{\Gamma}}_\text{in}$.
While at $\omega_\text{out} = \SI{186}{\tera\hertz}$, the effective input mode mostly consists of contribution from the first input mode $u_1(\Omega)$, going to $\omega_\text{out} = \SI{195}{\tera\hertz}$ exchanges the roles of $\hat{x}_{u_1}$ and $\hat{x}_{u_2}$ while the $p$-quadrature stays almost unaffected.

\begin{figure}
	\centering
	\includegraphics[width=0.45\textwidth]{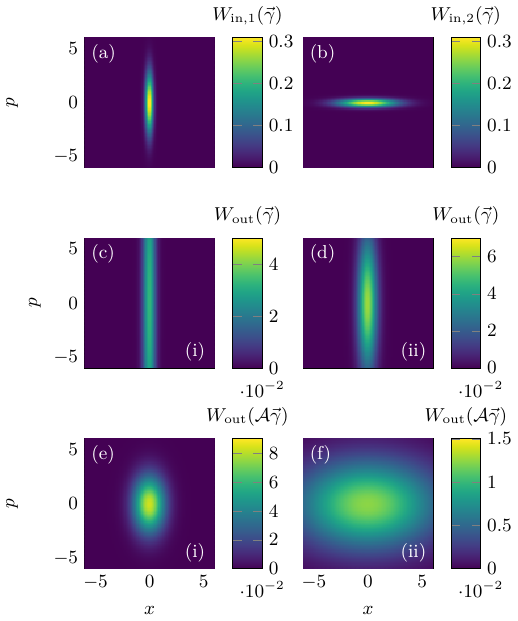}
	\caption{
		(a), (b) The two-mode squeezed input Wigner function of the first mode $u_1(\Omega)$ in (a) with a squeezing parameter of $r_1 = -1$ and of the second mode $u_2(\Omega)$ in (b) with a squeezing parameter of $r_2 = 1$.
		The covariance matrix of the input state is defined in Eq.~\eqref{eq:cov_sq}.
		(c), (d) The output Wigner function for a output mode $v(\omega)$ centered at $\omega_\text{out} / (2\pi) = \SI{186}{\tera\hertz}$ or $\omega_\text{out} / (2\pi) = \SI{195}{\tera\hertz}$, respectively.
		(e), (f) The same as (c), (d), however, the output Wigner function is transformed by $\mathcal{A}$ and shows the thermalization of the output Wigner function at $\omega_\text{out} = \SI{195}{\tera\hertz}$, as described in Sec.~\ref{ssec:squeezed}.
	}
	\label{fig:wigner_squeezed}
\end{figure}

The impact of the different effective input modes on the output Wigner function is best exemplified by a two-mode squeezed input state with covariance matrix
\begin{equation}\label{eq:cov_sq}
	\cov_\text{in} = \diag(e^{-2r}, e^{2r}, e^{2r}, e^{-2r})
,\end{equation}
expressed in the basis of $\hat{\vec{\Gamma}}_\text{in}$.
An example of the Wigner function with the covariance matrix of Eq.~\eqref{eq:cov_sq} and $r=1$ can be seen in Fig.~\ref{fig:wigner_squeezed} (a) for the reduced state of the phase space $(\hat{x}_{u_1}, \hat{p}_{u_1})$ and in (b) reduced to $(\hat{x}_{u_2}, \hat{p}_{u_2})$.

The output Wigner function for the parameter set (i) is shown in Fig.~\ref{fig:wigner_squeezed} (c) and for the parameter set (ii) in (d).
While it is not straight forward to see that the Wigner function for (i) and (ii) are thermal, i.e.  spread over a larger phase-space volume, the difference in thermalization of the two output quantum state becomes apparent if the Wigner function is transformed by $\mathcal{A}$, as shown in Fig.~\ref{fig:wigner_squeezed} (e) and (f).
The transformation with $\mathcal{A}$ mostly undoes the squeezing due to the nonlinear interaction, thus the covariance matrix of the rescaled output state is
\begin{equation}\label{eq:rescaled_cov}
	\cov_{\text{out},\mathcal{A}} = \cov_{\text{in,s}} + \mathcal{A}^{-1} \cov_B [\mathcal{A}^{-1}]^\T
.\end{equation}
Since the transformation with $\mathcal{A}$ is only one part of an open-systems dynamics, the state corresponding to the covariance matrix in Eq.~\eqref{eq:rescaled_cov} might not be physical.

The increased thermalization of the output quantum state, after transformation with $\mathcal{A}$, at (ii) compared to (i) can be explained as follows.
In case (i) the effective input quadrature, $(\hat{x}_\text{eff},\hat{p}_\text{eff})$, mostly correspond to the quadrature of the first input mode, $(\hat{x}_{u_1}, \hat{p}_{u_1})$, as described above and visible in \ref{fig:overlap} (a).
Therefore, the $x$-quadrature is squeezed and the $p$-quadrature is anti-squeezed.
In the case (ii), the $x$-quadratures of the input mode switch roles and $(\hat{x}_\text{eff},\hat{p}_\text{eff})$ correspond to $(\hat{x}_{u_2}, \hat{p}_{u_1})$.
Thus, both quadratures are anti-squeezed, leading to the increased phase-space volume.
The enlarged phase space volume will lead to an increase of the von Neumann entropy of the first mode, which quantifies the entanglement between the two modes in $S_R \hat{\vec{\Gamma}}_\text{in}$ of the (bipartite and pure) two-mode squeezed state.
Since the von Neumann entropy is increased by tracing over a mode, the process is called thermalization.
For a Gaussian quantum state, the von Neumann entropy can be calculated according to \cite{Holevo1999,Adesso2014,Weedbrook2012}.
There can be two origins for the thermalization.
First, an input state dependent part originating from entanglement breakage in the transformed input state when reducing the input modes to some fewer effective input modes.
Second, a state independent contribution, which can be understood as breakage of entanglement which was created in the nonlinear interaction.

The von Neumann entropy $S(\cov_{\text{in,s}})$ of the state with covariance matrix $\cov_{\text{in,s}} = P_R S_R \cov_\text{in} S_R^\T P_R^\T$ as a function of the central frequency $\omega_\text{out}$ of the output mode is shown in Fig.~\ref{fig:overlap} (b) and exhibits a clear increase of the entropy after $S_R$ exchanges the role of $\hat{x}_{u_1}$ and $\hat{x}_{u_2}$ as described above and shown in Fig.~\ref{fig:overlap} (a).
Entanglement generated by $S_R$ could explain the spike in the von Neumann entropy at $\omega_\text{out} \approx \SI{191}{\tera\hertz}$, since this feature is visible even with the vacuum as the input state and corresponds to the zero in $\mathcal{A}$, as can be seen in Fig.~\ref{fig:twoInOneOut} (a) and (b).
Another striking feature of the von Neumann entropy of the transformed input quantum state, is the minimum at $\omega_\text{out} \approx \SI{186}{\tera\hertz}$.
Here the effective quadratures most closely resemble the input quadratures and thus the reduced state is pure, which is why the von Neumann entropy goes to zero.

The von Neumann entropy $S(\cov_\text{out})$ of the actual output quantum state with covariance matrix $\cov_\text{out}$, as in Eq.~\eqref{eq:gaussian_output}, shows excess thermalization in the squeezing regime, i.e. for $\omega_\text{out} < \SI{186}{\tera\hertz}$, compared to the state with $\cov_{\text{in,s}}$.
This could be explained by the enhanced entanglement generation in the nonlinear interaction as compared to the beam-splitter regime.
After rescaling the output Wigner function with $\mathcal{A}$, resulting in the covariance matrix $\cov_{\text{out},A}$ of Eq.~\eqref{eq:rescaled_cov}, the von Neumann entropy $S(\cov_{\text{out},A})$ as a function of the central frequency of the output mode function mostly follows the entropy of the state with $\cov_{\text{in,s}}$, as shown in Fig.~\ref{fig:overlap} (b).
Therefore, most of the state independent thermalization is contained in $\mathcal{A}$, as the comparison between the von Neumann entropy of $\cov_\text{out}$ and $\cov_{\text{out},A}$ in Fig.~\ref{fig:overlap} (b) illustrates.
The entropy $S(\cov_{\text{out},A})$ exhibits an additional peak at $\omega_\text{out} \approx \SI{204}{\tera\hertz}$, which is due to the missing up-conversion at this parameter combination (as explained in Sec.~\ref{ssec:fock} for the Fock states), resulting in a singular $\mathcal{A}$.

Thus, the generalized Bloch-Messiah decomposition allows us to find an input-output relation between a different number of input and output modes.
By doing so, we can investigate the emergence of thermalization in a parametric up-conversion and find suitable modes to achieve the conversion of desired states.

\section{Conclusions}\label{sec:conclusion}
Nonlinear optics is key for the development of multimode (ultrafast) quantum optics.
In this work, we present the phase-space input-output relation of multimode Wigner functions representing pulsed quantum states undergoing a $\chi^{(2)}$ nonlinear interaction.
While the input modes are determined by the pulsed quantum state, the output modes are arbitrary and could, for example, be determined by a homodyne detection after the nonlinear interaction.
As the number of input (broadband) modes might not match the number of output modes, we introduce the generalized Bloch-Messiah decomposition (GBMD), based on a symplectic version of the singular value decomposition, to reduce the number of modes such that they coincide.
With the GBMD we show that the output Wigner function is related to the Wigner function of the reduced input state, convolved with a Gaussian phase-space function and rescaled by a non-symplectic matrix.

We elaborate on this result using two examples.
First, we consider a Fock state in a single THz-broadband mode up-converted to a single broadband mode in the optical frequencies.
We identify two regimes of the up-conversion determined by the central frequency of the output mode, which we call beam-splitter and squeezing regime, and characterize the effect of the different regimes on the output Wigner function.
Second, we describe the up-conversion of a squeezed pulse in two THz-broadband modes to a single optical pulse and show that the reduction of the input modes due to the GBMD can lead to entanglement breakage in the two-mode squeezed input state, resulting in thermalization of the output state.
We quantify the thermalization using the von Neumann entropy and identify a state independent thermalization due to the breakage of entanglement generated in the nonlinear interaction itself.

With the tools presented in this work, we are able to efficiently describe the nonlinear interactions of pulsed quantum states, necessary to advance quantum optics to the ultrafast regime, with possible applications spanning from optical quantum information technologies to efficient fast photo detection.

\begin{acknowledgments}
E. Hubenschmid acknowledges funding by the Deutsche Forschungsgemeinschaft (DFG) - Project No. 425217212 - SFB 1432.

V. Rueskov Christiansen acknowledges support from the Danish National Research Foundation through the Center of Excellence for Complex Quantum Systems (Grant agreement No. DNRF152).
\end{acknowledgments}

\section*{DATA AVAILABILITY}

The data and code that support the findings of this article are openly available \cite{SubcycleQ}.

\appendix

\begin{widetext}

\section{Input-Output relation of the quadrature operators}
In this Section we derive the quadrature input-output relation of Eq.~\eqref{eq:in_out_quadratures}.
\subsection{Orthogonalization procedure}\label{a:in_out_relation}
Applying the Gram-Schmidt orthogonalization procedure to $\{u_1, \ldots, u_N, g_1, \ldots g_M, f_1, \ldots f_M\}$ results in a set $\{u_1, \ldots u_N, h_1, \ldots h_{2M}\}$ of orthogonal mode functions with
\begin{align}
    h_i(\Omega) = \begin{cases}
	    \frac{1}{x_{i}}\left[g_{i}(\Omega) - \sum_{j=1}^{N} \inprod{u_j}{g_{i}}u_j(\Omega) -  \sum_{j=1}^{i-1} \inprod{h_j}{g_{i}}h_j(\Omega)\right] & \text{for } 1 \leq i \leq M \\
	    \frac{1}{z_{i-M}}\left[f_{i-M}(\Omega) - \sum_{j=1}^{N} \inprod{u_j}{f_{i-M}}u_j(\Omega) - \sum_{j=1}^{i-1} \inprod{h_j}{f_{i-M}}h_j(\Omega)\right] & \text{for } M < i \leq 2M
    \end{cases}
.\end{align}
If we solve the above equations for $g_i(\omega)$ and $f_i(\omega)$, we obtain
\begin{align}
	g_i(\Omega) &= x_i h_i(\Omega) + \sum_{j=1}^{N} \inprod{u_j}{g_{i}}u_j(\Omega) +  \sum_{j=1}^{i-1} \inprod{h_j}{g_{i}}h_j(\Omega), \\
	f_i(\Omega) &= z_i h_{i+M}(\Omega) + \sum_{j=1}^{N} \inprod{u_j}{f_{i}}u_j(\Omega) + \sum_{j=1}^{i+M-1} \inprod{h_j}{f_{i}}h_j(\Omega)
,\end{align}
which we can insert into the output operators of Eq.~\eqref{eq:output_operator},
\begin{align}
    \b_{v_i} &= \a_{f_i} + \a_{g_i}^\dagger \nonumber \\
	&= z_i \a_{h_{i+M}} + \sum_{j=1}^{N} \inprod{u_j}{f_{i}}^\ast\a_{u_j} + \sum_{j=1}^{i+M-1} \inprod{h_j}{f_{i}}^\ast\a_{h_j} 
	+ x_i \a_{h_i}^\dagger + \sum_{j=1}^{N} \inprod{u_j}{g_{i}}\a_{u_j}^\dagger +  \sum_{j=1}^{i-1} \inprod{h_j}{g_{i}} \a_{h_j}^\dagger \label{eq:output_quad_ap}
.\end{align}
By collecting the input, output and Gram-Schmidt operators into the vectors
\begin{align*}
    \hat{\vec{b}}_\text{out} &= (\hat{b}_{v_1}, \ldots, \hat{b}_{v_M}, \hat{b}_{v_1}^\dagger, \ldots, \hat{b}_{v_M}^\dagger)^\T, \\
    \hat{\vec{a}}_\text{in} &= (\hat{a}_{u_1}, \ldots, \hat{a}_{u_N}, \hat{a}_{u_1}^\dagger, \ldots, \hat{a}_{u_N}^\dagger)^\T, \\
    \hat{\vec{a}}_\perp &= (\hat{a}_{h_1}, \ldots, \hat{a}_{h_{2M}}, \hat{a}_{h_1}^\dagger, \ldots, \hat{a}_{h_{2M}}^\dagger)^\T, 
\end{align*}
and defining the matrix elements $[F_\text{in}]_{ji} = \inprod{u_i}{f_j}^\ast = \iint v_j^\ast(\omega) F(\omega,\Omega)u_i(\Omega)\dd \omega \dd \Omega$ and $[G_\text{in}]_{ji} = \inprod{u_i}{g_j} = \iint v_j^\ast(\omega) G^\ast(\omega,\Omega) u_i^\ast(\Omega) \dd \omega \dd \Omega$, which correspond to the matrices
\begin{align}
	F_\text{in} &= \begin{pmatrix}
		\inprod{u_1}{f_1}^\ast & \ldots & \inprod{u_N}{f_1}^\ast \\
		\vdots & & \vdots \\
		\inprod{u_1}{f_M}^\ast & \ldots & \inprod{u_N}{f_M}^\ast \\
	\end{pmatrix}, \quad
	G_\text{in} = \begin{pmatrix}
		\inprod{u_1}{g_1} & \ldots & \inprod{u_N}{g_1} \\
		\vdots & & \vdots \\
		\inprod{u_1}{g_M} & \ldots & \inprod{u_N}{g_M} \\
	\end{pmatrix}
,\end{align}
as well as defining the matrices
\begin{align}
	F_\perp &= \begin{pmatrix}
		\inprod{h_1}{f_1}^\ast & \ldots & \inprod{h_M}{f_1}^\ast & z_1 & 0 & 0 & \ldots & 0 \\
		\inprod{h_1}{f_2}^\ast & \ldots & \inprod{h_M}{f_2}^\ast & \inprod{h_{M+1}}{f_2}^\ast & z_2 & 0 & \ldots & 0 \\
		\vdots & \ddots &  & \vdots &  &  & \\
		\inprod{h_1}{f_M}^\ast & \ldots & \inprod{h_M}{f_M}^\ast & \ldots &  & \inprod{h_{2M-1}}{f_M}^\ast & z_M \\
	\end{pmatrix}, \\
	G_\perp &= \begin{pmatrix}
		x_1 & 0 & 0 & \ldots & & 0 & \ldots \\
		\inprod{h_1}{g_2} & x_2 & 0 & \ldots & & 0 & \ldots \\
		\vdots & \ddots & \vdots & & & &  \\
		\inprod{h_1}{g_M} & \ldots & & \inprod{h_{M-1}}{g_M} & x_M & 0 & \ldots   \\
	\end{pmatrix}
,\end{align}
we can utilize
\begin{align}
	A_\text{c} = \begin{pmatrix} 
	F_\text{in} & G_\text{in} \\
	G_\text{in}^\ast & F_\text{in}^\ast
\end{pmatrix}, \quad
	B_\text{c} = \begin{pmatrix} 
	F_\perp & G_\perp \\
	G_\perp^\ast & G_\perp^\ast
\end{pmatrix}
\end{align}
to recast Eq.~\eqref{eq:output_quad_ap} into $\hat{\vec{b}}_\text{out} = A_\text{c}\hat{\vec{a}}_\text{in} + B_\text{c}\hat{\vec{a}}_\perp$.
To go from the complex mode space to the real phase space, we can use the matrix
\begin{equation}
T = \frac{1}{\sqrt{2}}\begin{pmatrix} 
	\mathbb{I} & \iu \mathbb{I} \\
	\mathbb{I} & -\iu \mathbb{I}
\end{pmatrix}
\end{equation}
to transform the vectors $\hat{\vec{\Gamma}}_\text{out} = T^\dagger \hat{\vec{b}}_\text{out}$, $\hat{\vec{\Gamma}}_\text{in} = T^\dagger \hat{\vec{a}}_\text{in}$, $\hat{\vec{\Gamma}}_\perp = T^\dagger \hat{\vec{a}}_\perp$ and the matrices
\begin{align}
	A =& T^\dagger A_\text{c} T = \begin{pmatrix}
		\Re(F_\text{in} + G_\text{in}) & \Im(-F_\text{in} + G_\text{in}) \\
		\Im(F_\text{in} + G_\text{in}) & \Re(F_\text{in} - G_\text{in})
	\end{pmatrix} = \begin{pmatrix}
		\Re(F_\text{in}) & \Im(-F_\text{in}) \\
		\Im(F_\text{in}) & \Re(F_\text{in})
	\end{pmatrix} + \begin{pmatrix}
		\Re(G_\text{in}) & \Im(G_\text{in}) \\
		\Im(G_\text{in}) & \Re(-G_\text{in})
	\end{pmatrix}, \\
	B =& T^\dagger B_\text{c} T = \begin{pmatrix}
		\Re(F_\perp + G_\perp) & \Im(-F_\perp + G_\perp) \\
		\Im(F_\perp + G_\perp) & \Re(F_\perp - G_\perp)
	\end{pmatrix} = \begin{pmatrix}
		\Re(F_\perp) & \Im(-F_\perp) \\
		\Im(F_\perp) & \Re(F_\perp)
	\end{pmatrix} + \begin{pmatrix}
		\Re(G_\perp) & \Im(G_\perp) \\
		\Im(G_\perp) & \Re(-G_\perp)
	\end{pmatrix}
,\end{align}
accordingly.
Thus, multiplying the equation for $\hat{\vec{b}}_\text{out}$ by $T$ and using $T T^\dagger = \mathbb{I}$, we arrive at the input-output relation of Eq.~\eqref{eq:in_out_quadratures}.

\subsection{Example $N=M=1$}
Let us start with the simplest example used in the main text: One input mode $u(\omega)$ and one output mode $v(\omega)$ or $M = N = 1$.
In this case the matrix connecting the input and the output quadratures is,
\begin{equation}
    A = \begin{pmatrix}
	    \Re(\inprod{u}{f + g}) & \Im(\inprod{u}{f + g}) \\
	    \Im(\inprod{u}{-f + g}) & \Re(\inprod{u}{f - g})
    \end{pmatrix}
.\end{equation}
The matrix $A$ is symplectic up to a scalar since $A^\T \Omega_2 A = \Omega_2 (|\inprod{u}{f}|^2 - |\inprod{u}{g}|^2)$.
With the matrix
\begin{equation}
    B = \begin{pmatrix}
	    \Re(\inprod{h_1}{f}) + x & z & -\Im(\inprod{h_1}{f}) & 0 \\
	    \Im(\inprod{h_1}{f}) & 0 & \Re(\inprod{h_1}{f}) - x & z
    \end{pmatrix}
,\end{equation}
the contribution due to the initially unoccupied modes is
\begin{align}
    \cov_B  &= 
    \begin{pmatrix}
	    (\Re(\inprod{h_1}{f} + x)^2 & 0\\
	    0 & (\Re(\inprod{h_1}{f} - x)^2
    \end{pmatrix} +\left[\Im(\inprod{h_1}{f})^2 + z^2\right] \begin{pmatrix}
        1 & 0\\
        0 & 1
    \end{pmatrix}
.\end{align}

\section{The first-order unitary for THz-to-optical frequency conversion}\label{a:conversion}
In this section we will derive the input-output relation of the quadrature operators used in the example of THz-to-optical frequency conversion as described in Sec.~\ref{sec:first_order_unitary}.
We start from the time evolution operator,
\begin{equation}\label{eq:nl_unitary_first_order_a}
	\hat{U}_{\text{NL}} \approx \exp(\frac{1}{2}\iint_0^\infty J[\Omega, \omega] \hat{a}_\omega \hat{a}_\Omega \dd \omega \dd \Omega - \hc + \iint_0^\infty K[\Omega, \omega] \hat{a}_\omega^\dagger \hat{a}_\Omega \dd \omega \dd \Omega)
,\end{equation}
which is expressed in terms of a squeezing interaction given by the kernel $J(\Omega,\omega)$ and beam-splitting interaction described by $K(\Omega, \omega)$.
The unitary time-evolution operator is derived in \cite{Hubenschmid2025} and is based on a first-order Magnus expansion \cite{Magnus1954}.
For the geometric arrangement of the classical pulse and nonlienar crystal often encounter in electro-optic sampling \cite{Moskalenko2015}, the kernels are given by \cite{Hubenschmid2025}
\begin{align}
	J(\Omega, \omega) &= 2\frac{1}{\hbar} (2\pi)^{3/2}\left(\frac{\hbar}{4\pi\varepsilon_0 c A}\right)^{3/2} \sqrt{\frac{\omega + \Omega}{n(\omega + \Omega)}} \alpha(\omega + \Omega) \sqrt{\frac{\omega\Omega}{n(\omega)n(\Omega)}} \widehat{\lambda}[\Delta k(\Omega, \omega)] \label{eq:squeezing_nl}, \\
	K(\Omega, \omega) &= 2\frac{1}{\hbar} (2\pi)^{3/2}\left(\frac{\hbar}{4\pi\varepsilon_0 c A}\right)^{3/2} \sqrt{\frac{\abs{\omega - \Omega}}{n(\omega + \Omega)}} \left[\alpha(\Omega - \omega) + \alpha^\ast(\omega - \Omega)\right] \sqrt{\frac{\omega\Omega}{n(\omega)n(\Omega)}} \widehat{\lambda}[\Delta k(-\Omega, \omega)] \label{eq:beam_splitting_nl}
,\end{align}
with $\widehat{\lambda}(k) = \lambda\frac{L}{\sqrt{2\pi}}\sinc(k L / 2)$ being the Fourier transform of the transversal profile of the crystal, the wave-vector mismatch $\Delta k(\Omega, \omega) = k_{\Omega + \omega} - k_{\Omega} - k_{\omega}$ and the coherent amplitude $\alpha(\Omega)$ of the classical pulse, which vanishes for negative $\Omega$, the refractive index $n(\omega)$ of zinc-telluride \cite{Marple1964}, and the beam-waist area $A =  \pi (\SI{3}{\micro\meter})^2$ as well as $\hbar$, $c$ and $\varepsilon_0$ being the reduced Planck constant, the speed of light in vacuum and the permittivity respectively.

The first-order unitary approach developed in \cite{Onoe2022, Hubenschmid2024} allows to calculate the output operator $\hat{a}_v$ in terms of input operators.
As a first step, we use the first order of the Baker-Campbell-Hausdorff-formula together with the definition $f_K(\Omega) = -\int K^\ast(\Omega, \omega) v(\omega) \dd \omega$, $f_J(\Omega) = \int J^\ast(\Omega, \omega) v^\ast(\omega) \dd \omega$ to find
\begin{align}\label{eq:bkh}
	\hat{U}_{\text{NL}}^\dagger \hat{a}_{v} \hat{U}_{\text{NL}} &= \hat{a}_{v} + \int_0^\infty f_K^\ast(\Omega) \hat{a}_\Omega \dd \Omega + \int_0^\infty f_J(\Omega) \hat{a}^\dagger_\Omega \dd \Omega \nonumber \\
	&= \hat{a}_{v} + \theta_K \hat{a}_{K} + \theta_J \hat{a}_{J}^\dagger
.\end{align}
We normalize the operators $\hat{a}_K$ and $\hat{a}_J$ by defined by $f_K(\Omega)$ and $f_J(\Omega)$ respectively by
\begin{equation}
	\theta_K = \left(\int \abs{f_K(\Omega)}^2 \dd \Omega\right)^{1/2}, \quad\theta_J = \left(\int \abs{f_J(\Omega)}^2 \dd \Omega\right)^{1/2}
.\end{equation}
By defining $\theta = \sqrt{|\theta_K^2 - \theta_J^2|}$, $\vartheta_J = \theta_J / \theta$ and $\vartheta_K = \theta_K / \theta$, we can find a unitary operator, which transforms the output mode operator as,
\begin{align}\label{eq:first_order_in_out}
	\hat{U}_{\text{NL},1} \hat{a}_{v} \hat{U}_{\text{NL},1}^\dagger &= \begin{cases}
		\cos(\theta) \hat{a}_{v} + \sin(\theta)\left[\vartheta_K\hat{a}_{K} + \vartheta_J\hat{a}_{J}^\dagger\right] & \theta_K > \theta_J \\
		\cosh(\theta) \hat{a}_{v} + \sinh(\theta)\left[\vartheta_K\hat{a}_{K} + \vartheta_J\hat{a}_{J}^\dagger\right] & \theta_K < \theta_J
	\end{cases}
,\end{align}
assuming the THz-frequencies don't overlap with the output mode, i.e., $[\hat{a}_v, \hat{a}_K^\dagger] = [\hat{a}_v, \hat{a}_J^\dagger] = 0$.
Since the commutator 
\begin{equation}
    [\vartheta_K \hat{a}_{K} + \vartheta_J \hat{a}_{J}^\dagger, \vartheta_K \hat{a}_{K}^\dagger + \vartheta_J \hat{a}_{J}] = \begin{cases}
        1 & \theta_K > \theta_J \\
        -1 & \theta_K < \theta_J
    \end{cases}
\end{equation}
corresponds to $\vartheta_K \hat{a}_{K} + \vartheta_J \hat{a}_{J}^\dagger$ being either a creation or annihilation operator, the unitary evolution in Eq.~\eqref{eq:first_order_in_out} has to correspond to either a beam splitting or squeezing interaction.
In both cases, the first order of a Taylor expansion around  $\theta = 0$ agrees with the result obtained from the Baker-Campbell-Hausdorff-formula in Eq.~\eqref{eq:bkh}.
Although the above argument only guarantees the validity of the first-order unitary for small $\theta$, calculations involving the second-order unitary suggest a large parameter range, the first-order unitary holds valid \cite{Onoe2022}.

By using $\mu$ and $\nu$ as in Eq.~\eqref{eq:mu_nu} and comparing the input-output relation from Eq.~\eqref{eq:first_order_in_out} and \eqref{eq:in_out_quadratures}, we find $F(\omega, \Omega) = \mu \delta(\omega - \Omega) - \frac{\nu}{\theta} K(\Omega, \omega)$ and $G(\Omega, \omega) = \frac{\nu}{\theta} J(\Omega, \omega)$, resulting in 
\begin{align}
	f(\Omega) &= \frac{1}{\zeta} \int F^\ast(\omega, \Omega) v(\omega) \dd \omega = \frac{1}{\zeta}\left[\mu v(\Omega) + \nu \vartheta_K f_K(\Omega) \right], \\
	\zeta^2 &= \mu^2 \inprod{v}{v} + 2\nu\mu \Re(\inprod{v}{f_K}) + \nu^2 \vartheta_K^2 \inprod{f_K}{f_K} = \mu^2 + \nu^2 \vartheta_K^2, \\
	g(\Omega) &= \frac{\nu}{\xi\theta} \int \dd \omega G^\ast(\omega, \Omega) v^\ast(\omega) =  \frac{\nu}{\xi} \vartheta_J f_J(\Omega), \\
	\xi^2 &= \nu^2 \vartheta_J^2 \inprod{f_J}{f_J} = \nu^2 \vartheta_J^2
.\end{align}

Therefore, the matrix elements connecting the input and output modes are,
\begin{align}
    [F_\text{in}]_{1i} &= \inprod{u_i}{f}^\ast = \frac{1}{\zeta}\left[\mu \inprod{u_i}{v}^\ast + \nu \vartheta_K \inprod{u_i}{f_K}^\ast\right] = \frac{\nu \vartheta_K}{\zeta} \inprod{u_i}{f_K}^\ast, \\
    [G_\text{in}]_{1i} &= \inprod{u_i}{g} = \frac{\nu}{\xi}\vartheta_J\inprod{u_i}{f_J}
.\end{align}

\section{Generalized Bloch Messiah decomposition}\label{a:gbmd}
In general it is not possible to decompose $A$ using a Bloch-Messiah decomposition, since the matrix is not symplectic.
For this reason we introduce the generalized Bloch-Messiah decomposition (GBMD) based on the symplectic singular value-like decomposition (SSVD) introduced by Xu \cite{Xu2003,Xu2005}.
In principle the matrix $A$ could not be full rank, if for example the overlap between one input mode and all modes $f_i(\omega)$ and $g_i(\omega)$ is zero or the other way around.
However, this transformation is trivial and will be excluded here.
We therefore assume $A \Omega A^\T$ is non-singular, i.e., $\rank(A \Omega A^\T) = \min(2M, 2N)$.

First, we consider the case with more input modes than output modes $N > M$.
The SSVD then states that there exist a decomposition
\begin{equation}\label{eq:rBMD}
    A = O_R \begin{pmatrix}
        \Sigma_{R} & 0 & 0 & 0 \\
        0 & 0 & \Sigma_{R} & 0
    \end{pmatrix}S_R = O_R \begin{pmatrix}
        \Sigma_{R} & 0 \\
        0 &\Sigma_{R}
    \end{pmatrix} \begin{pmatrix}
        \mathbb{I}_M & 0 & 0 & 0 \\
        0 & 0 & \mathbb{I}_M & 0
    \end{pmatrix}S = \mathcal{A}P_R S_R
\end{equation}
with $O_R \in \mathbb{R}^{2M \times 2M}$ orthogonal, $\mathcal{A} = O_R\text{diag}(\Sigma_{R}, \Sigma_{R}) \in \mathbb{R}^{2M \times 2M}$ regular, $\Sigma_{R}$ positive diagonal, $S_R \in \mathbb{R}^{2N \times 2N}$ symplectic and $P_R$ projecting from $2N$ dimensions to $2M$ dimensions.
By defining $\hat{\Vec{\Gamma}}_{\text{in,s}} = P_R S_R \hat{\Vec{\Gamma}}_{\text{in}} = (\hat{x}_{\text{eff},1}, \ldots, \hat{x}_{\text{eff},M}, \hat{p}_{\text{eff},1}, \ldots, \hat{p}_{\text{eff},M})^\T$ the input-output relation \eqref{eq:in_out_quadratures} can be recast into
\begin{equation}
	\hat{\Vec{\Gamma}}_{\text{out}} =  \mathcal{A} \hat{\Vec{\Gamma}}_{\text{in,s}} + B\vec{\Gamma}_\perp
,\end{equation}
which now only includes the first $2M$ quadrature operators and the orthogonal modes in $\vec{\Gamma}_\perp$.

Second, we assume more output modes than input modes, i.e., $M > N$, and decompose
\begin{equation}\label{eq:lBMD}
    A^\T = O_L \begin{pmatrix}
        \Sigma_{L} & 0 & 0 & 0 \\
        0 & 0 & \Sigma_{L} & 0
    \end{pmatrix}S_L = O_L \begin{pmatrix}
        \Sigma_{L} & 0 \\
        0 &\Sigma_{L}
    \end{pmatrix} \begin{pmatrix}
        \mathbb{I}_N & 0 & 0 & 0 \\
        0 & 0 & \mathbb{I}_N & 0
    \end{pmatrix}S_L = \mathcal{A}^\T P_L S_L
\end{equation}
with $O_L \in \mathbb{R}^{2N \times 2N}$ orthogonal, $\mathcal{A} = O_L\text{diag}(\Sigma_{L}, \Sigma_{L}) \in \mathbb{R}^{2N \times 2N}$ regular, $\Sigma_{L}$ positive diagonal, $S_L \in \mathbb{R}^{2M \times 2M}$ symplectic and $P_L$ projecting from $2M$ dimensions to $2N$ dimensions.

Together we can write:
\begin{equation}\label{eq:gbmd_a}
    A = S_L^\T P_L^\T \mathcal{A} P_R S_R
\end{equation}
with either $S_L = P_L = \mathbb{I}$ for $N>M$, $S_R = P_R = \mathbb{I}$ for $M>N$ or $S_L = P_L = S_R = P_R = \mathbb{I}$ if $N = M$.
In all three cases, $\mathcal{A} \in \mathbb{R}^{2k \times 2k}$ with $k = \min(M,N)$ is regular and we call Eq.~\eqref{eq:rBMD} and \eqref{eq:lBMD} the generalized Bloch-Messiah decompositions.
An algorithm to compute the SSVD can be found in \cite{Xu2005} and in the next section we will present an example of the GBMD for one output mode and $N$ input modes.
In the version we present here, $\Sigma_R$ might not be positive definite, which does not matter for our purposes here, since $\Sigma$ will be absorbed into $\mathcal{A}$ anyways.

\subsection{Example of the generalized Bloch-Messiah decomposition with one output mode and $N$ input modes}\label{sm:example_gbmd}\label{a:gbmd_example}
If we assume one output mode, $M=1$, and multiple input modes, $N>1$, the matrix $A$ generally takes the form
\begin{equation}
	A = \begin{pmatrix} 
		\begin{array}{cccc|cccc}
			\bullet & \bullet & \bullet & \ldots & \bullet & \bullet & \bullet & \ldots \\
			\hline
			\bullet & \bullet & \bullet & \ldots & \bullet & \bullet & \bullet & \ldots
		\end{array}
	\end{pmatrix}
,\end{equation}
with $\bullet$ marking the (possibly nonzero) entries of the matrix.
Folowing Xu \cite{Xu2005}, we can use the symplectic Housholder transformations $U_1, U_2$, $U = U_2 U_1$ acting on the second row of $A$, and symplectic Givens rotation $O_R$ (in the order $U_1, O_R, U_2$) to transform the matrix $A$.
First, $U_1^\T$ eliminates $A_{22}$ to $A_{2N}$, than $O_R^\T$ elimates $A_{21}$ and lastly $U_2^\T$ can be used to eliminate $A_{2(N+1)}$ to $A_{2(2N)}$, since $A_{21}$ to $A_{2N}$ is already zero, resulting in

\begin{equation}\label{eq:tranformed_A}
	O_R^\T A U = \begin{pmatrix} 
		\begin{array}{cccc|cccc}
			\bullet & \bullet & \bullet & \ldots & \bullet & \bullet & \bullet & \ldots \\
			\hline
			0 & 0 & 0 & \ldots & \bullet & 0 & 0 & \ldots
		\end{array}
	\end{pmatrix} = \begin{pmatrix} 
		\begin{array}{cc|cc}
			R_{11} & R_{12} & R_{13} & R_{14} \\
			\hline
			0 & 0 & R_{23} & 0
		\end{array}
	\end{pmatrix}
.\end{equation}
Since $R_{23} R_{11}$ might be negative, one can follow \cite{Xu2005} or in this case we can use $s = \sign(R_{23} R_{11})$ and multiply Eq.~\eqref{eq:tranformed_A} from the left with $\begin{pmatrix}- s & 0 \\ 0 & 1\end{pmatrix}$, which will later be absorbed into $\mathcal{A}$.
With $R_{23} R_{11}$ being positive, we can define the symplectic matrix
\begin{equation}
	S_R^{-1} = U \begin{pmatrix} 
		\sqrt{R_{23}/R_{11}} & -R_{12}/R_{11} & - R_{13}^\T/\sqrt{R_{11}R_{23}} & -R_{14}/R_{11} \\
		0 & \mathbb{I} & - R_{14}^\T/\sqrt{R_{11}R_{23}} & 0 \\
		0 & 0 & \sqrt{R_{11}/R_{23}} & 0 \\
		0 & 0 & -R_{12}^\T/\sqrt{R_{11}R_{23}} & \mathbb{I}
	\end{pmatrix}
.\end{equation}
Using $\Sigma_R = \sqrt{R_{23}R_{11}}$ we obtain the GBMD
\begin{equation}
	A = O_R\begin{pmatrix} 
	\begin{array}{ccc|ccc}
		s\Sigma_R & 0 & \ldots & 0 & 0 & \ldots \\
			\hline
		0 & 0 & \ldots & \Sigma_R & 0 & \ldots
		\end{array}
	\end{pmatrix}  S_R = \mathcal{A} P_R  S_R
,\end{equation}
with $\mathcal{A} = O_R \begin{pmatrix}
    s \Sigma_R & 0 \\
    0 & \Sigma_R
\end{pmatrix}$ and $P_R = \begin{pmatrix} 
	\begin{array}{ccc|ccc}
		1 & 0 & \ldots & 0 & 0 & \ldots \\
			\hline
		0 & 0 & \ldots & 1 & 0 & \ldots
		\end{array}
	\end{pmatrix}$.
It should be mentioned, that the phase-space basis obtained from $P_R S_R \hat{\vec{\Gamma}}_\text{in}$, might be squeezed.
To undo the squeezing, we can calculate the effective squeezing parameter using $r_\text{eff} = \ln(\sqrt{\sum_{i=1}^N [S_R]_{1j}}/\sqrt{\sum_{i=1}^N [S_R]_{3j}})$ and multiply $S_R$ from the left and $\mathcal{A}$ from the left by the respective squeezing or anti squeezing of the first mode.

\section{Quantum state input-output relation in phase space}\label{a:inOutPhaseSpace}
To calculate the Wigner function of the output state we use the charactersitic function of the initial quantum state $\hat{\rho}$ with some modes occupied by the quantum state and the rest in a vacuum $\hat{\rho} = \hat{\rho}_\text{in} \otimes \ket{0}_\perp \bra{0}$. Using Eq.~\eqref{eq:in_out_quadratures} and Eq.~\eqref{eq:gbmd_a} the characteristic function can be rewritten as
\begin{align}
    \chi_\text{out}(\vec{\beta}) &= \tr[\hat{\rho} \exp(-\iu \hat{\vec{\Gamma}}_\text{out}^\T \Omega_M \vec{\beta})] \nonumber \\
	&= \tr[\hat{\rho} \exp(-\iu [A\hat{\vec{\Gamma}}_\text{in} + B\hat{\vec{\Gamma}}_\perp]^\T \Omega_M \vec{\beta})] \nonumber \\
	&= \tr[\hat{\rho}_\text{in} \exp(-\iu [A\hat{\vec{\Gamma}}_\text{in}]^\T\Omega_M \vec{\beta})] \tr[\ket{0}_\perp \bra{0}\exp( -\iu[B\hat{\vec{\Gamma}}_\perp]^\T \Omega_M \vec{\beta})] \nonumber \\
	&= \tr[\hat{\rho}_\text{in} \exp(-\iu [P_R S_R \hat{\vec{\Gamma}}_\text{in}]^\T \mathcal{A}^\T \Omega_M P_L S_L^{-1}\vec{\beta})] \tr[\ket{0}_\perp \bra{0}\exp( - \iu \hat{\vec{\Gamma}}_\perp^\T B^\T \Omega_M \vec{\beta})]
.\end{align}
Therefore, the generalized Bloch Messiah decomposition in Eq.~\eqref{eq:gbmd} allows us to find a mode basis for which we can use the reduced state $\hat{\rho}_{\text{in,s}} = \tr_r(\hat{\rho}_\text{in})$ of the $k = \min\{N,M\}$ modes and the trace over the last $N-k$ modes in $S_R \hat{\Vec{\Gamma}}_{\text{in}}$.
Thus, only the modes in $\hat{\Vec{\Gamma}}_{\text{in,s}} = P_R S_R \hat{\Vec{\Gamma}}_{\text{in}}$ are relevant to the output quantum state.
Furthermore, we define $\vec{\beta}_\text{s} = P_LS_L^{-1} \vec{\beta}$ and $\vec{\beta}_\text{e}$ containing the $M-k$ modes in S$_L^{-1} \vec{\beta}$, but not in $\vec{\beta}_\text{s}$.
Thus, we have
\begin{align}
	\chi_\text{out}(\vec{\beta}) &= \tr[\hat{\rho}_{\text{in,s}} \exp(-\iu \hat{\vec{\Gamma}}_{\text{in,s}}^\T \mathcal{A}^\T \Omega_L \vec{\beta}_\text{s})] \tr[\ket{0}_\perp \bra{0}\exp( - \iu \hat{\vec{\Gamma}}_\perp^\T B^\T \Omega_M \vec{\beta})]
.\end{align}
From here on we will drop the index of $\Omega$, the dimension of which is implied by the context.
By defining $\Tilde{\mathcal{A}} = \Omega^\T \mathcal{A}^\T \Omega$ ($\Tilde{\mathcal{A}} = \mathcal{A}^{-1}$ if $\mathcal{A}$ is symplectic) and $\Tilde{B} = \Omega^\T B^\T \Omega$ we can write
\begin{align}
	\chi_\text{out}(\vec{\beta}) &= \tr[\hat{\rho}_{\text{in,s}} \exp(-\iu \hat{\vec{\Gamma}}_{\text{in,s}}^\T \Omega \Tilde{\mathcal{A}} \vec{\beta}_\text{s})]\tr[\ket{0}_\perp \bra{0}\exp(\hat{\vec{\Gamma}}_\perp^\T \Omega\Tilde{B} \vec{\beta})] \nonumber \\
	&= \chi_{\text{in,s}}(\Tilde{\mathcal{A}}\vec{\beta}_\text{s})\chi_\text{vac}(\Tilde{B}\vec{\beta})
,\end{align}
with
\begin{align}
    \chi_\text{vac}(\Tilde{B}\vec{\beta}) &= \exp(-\frac{1}{4} [\Tilde{B}\vec{\beta}]^\T \Tilde{B}\vec{\beta}) \nonumber \\
    &= \exp(-\frac{1}{4} \vec{\beta}^\T \Omega^\T B \Omega \Omega^\T B^\T \Omega \vec{\beta}) \nonumber \\
    &= \exp(-\frac{1}{4} \vec{\beta}^\T \Omega \cov_B \Omega^\T \vec{\beta}) \nonumber \\
	&= \chi_B(\vec{\beta})
\end{align}
and $\cov_B = B B^\T$.
We decompose $\vec{\gamma}^\T \Omega \vec{\beta} = \vec{\gamma}_\text{e}^\T \Omega \vec{\beta}_\text{e} + \vec{\gamma}_\text{s}^\T \Omega \vec{\beta}_\text{s}$ and $\vec{\beta}^\T \Omega \cov_B \Omega^\T \vec{\beta} = \vec{\beta}_\text{s}^\T\Omega\cov_{\text{th,ss}}\Omega^\T\vec{\beta}_\text{s} + \vec{\beta}_\text{e}^\T\Omega\cov_{\text{th,es}}\Omega^\T\vec{\beta}_\text{s} + \vec{\beta}_\text{s}^\T\Omega\cov_{\text{th,se}}\Omega^\T\vec{\beta}_\text{e} + \vec{\beta}_\text{e}^\T\Omega\cov_{\text{th,ee}}\Omega^\T\vec{\beta}_\text{e}$ and using that $\cov_{B}$ is symmetric we can solve,
\begin{align}
	&\int_{\mathbb{R}^{2(M-L)}} \exp(\iu \vec{\gamma}_\text{e}^\T \Omega \vec{\beta}_\text{e}-\frac{1}{4}[\vec{\beta}_\text{e}^\T\Omega\cov_{\text{th,es}}\Omega^\T\vec{\beta}_\text{s} + \vec{\beta}_\text{s}^\T\Omega\cov_{\text{th,se}}\Omega^\T\vec{\beta}_\text{e} + \vec{\beta}_\text{e}^\T\Omega\cov_{\text{th,ee}}\Omega^\T\vec{\beta}_\text{e}]) \dd^{2(M-L)} \beta_\text{e} \nonumber\\
	&=\sqrt{\frac{(4\pi)^{2(M-L)}}{\det(\cov_{\text{th,ee}})}}\exp(
	\frac{1}{4}\vec{\beta}_\text{s}^\T \Omega\cov_{\text{th,se}}\cov_{\text{th,ee}}^{-1}\cov_{\text{th,es}}\Omega^\T\vec{\beta}_\text{s}
	+\iu \vec{\gamma}_\text{e}^\T \cov_{\text{th,ee}}^{-1}\cov_{\text{th,es}}\Omega^\T\vec{\beta}_\text{s}
	-\vec{\gamma}_\text{e}^\T \Omega\cov_{\text{th,ee}}^{-1}\vec{\gamma}_\text{e})
.\end{align}
This motivates the definition $\cov_{\text{schur}} = \cov_{\text{th,ss}} - \cov_{\text{th,se}} \cov_{\text{th,ee}}^{-1}\cov_{\text{th,es}}$ and $\vec{\gamma}_\text{schur} = \cov_{\text{th,es}}\cov_{\text{th,ee}}^{-1}\vec{\gamma}_\text{e}$.
Since $\Tilde{\mathcal{A}}$ is regular and invertible, such that $\Tilde{\mathcal{A}}^{-1} = \Omega^\T (\mathcal{A}^{-1})^\T \Omega$, we can calculate the Wigner function of the amplified state,
	\begin{align}
		W_\text{out}(\vec{\gamma}) &= \frac{1}{(2\pi)^{2M}} \int_{\mathbb{R}^{2M}} \exp(\iu \vec{\gamma}^\T \Omega \vec{\beta}) \chi_\text{out}(\vec{\beta}) \dd^{2M} \beta \nonumber \\
		&= \frac{1}{(2\pi)^{2M}} \int_{\mathbb{R}^{2M}} \exp(\iu \vec{\gamma}^\T \Omega \vec{\beta} - \frac{1}{4} \vec{\beta}^\T \Omega \cov_B \Omega^\T \vec{\beta}) \chi_{\text{in,s}}(\Tilde{\mathcal{A}}\vec{\beta}) \dd^{2M} \beta \nonumber \\
		&= \left[(2\pi)^{2(M+L)}\det(\cov_{\text{th,ee}}/2)\right]^{-\frac{1}{2}} \int_{\mathbb{R}^{2L}} \exp(\iu [\vec{\gamma}_\text{s} + \vec{\gamma}_\text{schur}]^\T \Omega \vec{\beta}_\text{s} - \frac{1}{4} \vec{\beta}_\text{s}^\T \Omega \cov_\text{schur} \Omega^\T\vec{\beta}_\text{s}) \chi_{\text{in,s}}(\Tilde{\mathcal{A}}\vec{\beta}_\text{s}) \dd^{2L} \beta_\text{s} \nonumber \\
		&= \left[(2\pi)^{2(M + L)}\det(\cov_{\text{th,ee}}/2)\right]^{-\frac{1}{2}} \nonumber\\
		&\qquad\times \int_{\mathbb{R}^{2L}} \exp(\iu [\vec{\gamma}_\text{s} + \vec{\gamma}_\text{schur}]^\T \Omega \Tilde{\mathcal{A}}^{-1}\Tilde{\mathcal{A}}\vec{\beta}_\text{s} - \frac{1}{4} [\Tilde{\mathcal{A}}^{-1}\Tilde{\mathcal{A}}\vec{\beta}_\text{s}]^\T \Omega \cov_\text{schur} \Omega^\T\Tilde{\mathcal{A}}^{-1}\Tilde{\mathcal{A}}\vec{\beta}_\text{s}) \chi_{\text{in,s}}(\Tilde{\mathcal{A}}\vec{\beta}_\text{s}) \dd^{2L} \beta_\text{s} \nonumber \\
		&= \frac{1}{\abs{\det(\mathcal{A})}}\left[(2\pi)^{2(M + L)}\det(\cov_{\text{th,ee}}/2)\right]^{-\frac{1}{2}} \nonumber\\
		&\qquad\times \int_{\mathbb{R}^{2L}} \exp(\iu [\vec{\gamma}_\text{s} + \vec{\gamma}_\text{schur}]^\T \Omega \Tilde{\mathcal{A}}^{-1}\vec{\beta}_\text{s} - \frac{1}{4} \vec{\beta}_\text{s}^\T[\Tilde{\mathcal{A}}^{-1}]^\T \Omega \cov_\text{schur} \Omega^\T\Tilde{\mathcal{A}}^{-1}\vec{\beta}_\text{s}) \chi_{\text{in,s}}(\vec{\beta}_\text{s}) \dd^{2L} \beta_\text{s} \nonumber \\
		&= \left[(2\pi)^{2(M + L)}\det(\cov_{\text{th,ee}}/2)\abs{\det(\mathcal{A})}^2\right]^{-\frac{1}{2}} \nonumber\\
		&\qquad\times \int_{\mathbb{R}^{2L}} \exp(\iu [\mathcal{A}^{-1}\{\vec{\gamma}_\text{s} + \vec{\gamma}_\text{schur}\}]^\T \Omega \vec{\beta}_\text{s} - \frac{1}{4} \vec{\beta}_\text{s}^\T\Omega\mathcal{A}^{-1} \cov_\text{schur} [\mathcal{A}^{-1}]^\T\Omega^\T\vec{\beta}_\text{s}) \chi_{\text{in,s}}(\vec{\beta}_\text{s}) \dd^{2L} \beta_\text{s}
.\end{align}
By defining
\begin{align}
	G(\vec{\gamma}_\text{s}) &= \left[(2\pi)^{2(M - L)}\det(\cov_{\text{th,ee}}/2)\abs{\det(\mathcal{A})}^2\right]^{-\frac{1}{2}} \int_{\mathbb{R}^{2L}} \exp(\iu \vec{\gamma}_\text{s}^\T \Omega \vec{\beta}_\text{s} - \frac{1}{4} \vec{\beta}_\text{s}^\T\Omega\mathcal{A}^{-1} \cov_\text{schur} [\mathcal{A}^{-1}]^\T\Omega^\T\vec{\beta}_\text{s}) \dd^{2L} \beta_\text{s} \nonumber\\
	&= 2^M\left[(2\pi)^{2(M - 2L)}\det(\cov_{\text{schur}})\det(\cov_{\text{th,ee}})\right]^{-\frac{1}{2}} \exp(-\vec{\gamma}_\text{s}^\T\mathcal{A} \cov_\text{schur}^{-1} \mathcal{A}^\T\vec{\gamma}_\text{s})
\end{align}
and
\begin{equation}
	W_{\text{in,s}}(\vec{\gamma}_\text{s}) = \frac{1}{(2\pi)^{2L}} \int_{\mathbb{R}^{2N}} \exp(\iu \vec{\gamma}_\text{s}^\T \Omega \vec{\beta}_\text{s}) \chi_{\text{in,s}}(\vec{\beta}_\text{s}) \dd^{2L} \beta_\text{s}
,\end{equation}
the output Wigner function is a smoothed out and transformed version of the input Wigner function,
\begin{equation}\label{eq:outputWignerfunction}
	W_\text{out}(\vec{\gamma}) = \Big(W_{\text{in,s}} \ast G\Big)(\mathcal{A}^{-1}[\vec{\gamma}_\text{s} + \vec{\gamma}_\text{schur}])
.\end{equation}

\subsection{Fock states}
Let us assume an $n$-photon Fock state in the first mode, then the characteristic function is
\begin{align}
    \exp(-\iu\vec{\hat{\Gamma}}_\text{in}\Omega_2\vec{\beta}) &= \exp(-\iu\beta_{x}\hat{p} + \iu\beta_{p}\hat{x}) \nonumber \\
    &= \exp(\frac{1}{\sqrt{2}}\beta_{x}(\hat{a}^\dagger - \hat{a}) + \frac{\iu}{\sqrt{2}}\beta_{p}(\hat{a}^\dagger + \hat{a})) \nonumber \\
    &= \exp(\frac{1}{\sqrt{2}}(\beta_{x} + \iu\beta_{p})\hat{a}^\dagger - \frac{1}{\sqrt{2}}(\beta_{x} - i\beta_{p})\hat{a})
,\end{align}
i.e., $\alpha = \frac{1}{\sqrt{2}}(\beta_{x} + \iu\beta_{p})$.
This allows us to apply the known formula for the characteristic function of the Fock state \cite{Vogel2006},
\begin{align}
    \chi_\text{in}(\vec{\beta}) &= \bra{n}\exp(\alpha \hat{a}^\dagger - \alpha^\ast \hat{a})\ket{n} \nonumber \\
    &= L_n(|\alpha|^2)e^{-|\alpha|^2/2} \nonumber \\
	&= L_n(||\vec{\beta}||^2 / 2)e^{-||\vec{\beta}||^2/4}
,\end{align}
and by inserting the charactersitic function of the Fock state into Eq.~\eqref{eq:outputWignerfunction}, we obtain the output Wigner function 
\begin{align}
    W_\text{out}(\vec{\gamma}) &= \frac{1}{(2\pi)^2\abs{\det(A)}} \int_{-\infty}^\infty \exp(-\frac{1}{4} \vec{\beta}^\T \Omega_2 A^{-1} \cov_B [A^{-1}]^\T \Omega_2^\T \vec{\beta} + \iu [A^{-1}\vec{\gamma}]^\T \Omega_2 \vec{\beta}) \chi_\text{in}(\vec{\beta}) d^4 \beta
.\end{align}
Using the SVD $A^{-1} \cov_B [A^{-1}]^\T = \sigma_x \vec{e}_x \vec{e}_x^\T + \sigma_p \vec{e}_p \vec{e}_p^\T$, defining $A^{-1}\vec{\gamma} = y\vec{e}_x + z\vec{e}_p$ and integrate over $\vec{\beta} = x \vec{e}_x + p \vec{e}_p$, we find the output Wigner function,
\begin{align}
	W_\text{out}(\vec{\gamma}) &= \frac{1}{(2\pi)^2\abs{\det(A)}} \iint_{-\infty}^\infty \exp(-\frac{1}{4} (\sigma_x + 1) x^2 -\frac{1}{4} (\sigma_p + 1) p^2 + \iu y p - \iu z x) L_n\left(\frac{x^2 + p^2}{2}\right) \dd x \dd p \nonumber \\
	&= \frac{\exp(-\frac{z^2}{\sigma_x + 1}-\frac{y^2}{\sigma_p + 1})}{\pi\abs{\det(A)}\sqrt{(\sigma_x + 1)(\sigma_p + 1)}} \sum_{m=1}^n \left(\frac{\sigma_x - 1}{\sigma_x + 1}\right)^{m}L^{(-\frac{1}{2})}_{m}\left(\frac{-2 z^2}{\sigma_x^2-1}\right) \left(\frac{\sigma_p - 1}{\sigma_p + 1}\right)^{n-m}L^{(-\frac{1}{2})}_{n-m}\left(\frac{-2 y^2}{\sigma_p^2-1}\right)
.\end{align}
If $\sigma_x = \sigma_p =: \sigma$ the output Wigner function corresponds to a ensemble of Fock states with different photon occupations,
\begin{align}
	W_\text{out}(\vec{\gamma}) &= \frac{4\pi}{(\sigma + 1)}e^{-\frac{z^2 + y^2}{\sigma + 1}}\frac{1}{(2\pi)^2\abs{\det(A)}} \left(\frac{\sigma - 1}{\sigma + 1}\right)^{n} \sum_{m=1}^n L^{(-\frac{1}{2})}_{m}\left(\frac{-2 z^2}{\sigma^2-1}\right) L^{(-\frac{1}{2})}_{n-m}\left(\frac{-2 y^2}{\sigma^2-1}\right) \nonumber \nonumber\\
	 &= \frac{1}{\pi(\sigma+1)\abs{\det(A)}} \left(\frac{\sigma-1}{\sigma+1}\right)^{n} L_{n}\left(-2\frac{||A^{-1}\vec{\gamma}||^2}{\sigma^2-1}\right)\exp\left(-\frac{||A^{-1}\vec{\gamma}||^2}{\sigma}\right)
.\end{align}

\subsection{Gaussian states}
For the case of Gaussian input state, the convolution in Eq.~\eqref{eq:outputWignerfunction} can be solved, since the convolution of two Gaussians is again a Gaussian.
Starting from the characteristic funtion of the reduced input state,
\begin{equation}
	\chi_{\text{in,s}}(\vec{\beta}) = \exp\left(-\frac{1}{4}\vec{\beta}^\T \Omega \cov_{\text{in,s}}\Omega^\T\vec{\beta}\right)
\end{equation}
we find the output Wigner function,
\begin{align}
	W_\text{out}(\vec{\gamma}) &= \left[(2\pi)^{2(M + L)}\det(\cov_{\text{th,ee}}/2)\abs{\det(\mathcal{A})}^2\right]^{-\frac{1}{2}} \nonumber \\
	&\qquad\times \int_{\mathbb{R}^{2L}} \exp(\iu [\mathcal{A}^{-1}\{\vec{\gamma}_\text{s} + \vec{\gamma}_\text{schur}\}]^\T \Omega \vec{\beta}_\text{s} - \frac{1}{4} \vec{\beta}_\text{s}^\T\Omega\{\mathcal{A}^{-1} \cov_\text{schur} [\mathcal{A}^{-1}]^\T + \cov_{\text{in,s}}\}\Omega^\T\vec{\beta}_\text{s}) \dd^{2L} \beta_\text{s} \nonumber \\
	&= \left[\pi^{2M}\det(\cov_{\text{th,ee}})\abs{\det(\mathcal{A})}^2 \det(\mathcal{A}^{-1} \cov_\text{schur} [\mathcal{A}^{-1}]^\T + \cov_{\text{in,s}})\right]^{-\frac{1}{2}} \nonumber \\
	&\qquad\times \exp(-[\mathcal{A}^{-1}\{\vec{\gamma}_\text{s} + \vec{\gamma}_\text{schur}\}]^\T \{\mathcal{A}^{-1} \cov_\text{schur} [\mathcal{A}^{-1}]^\T + \cov_{\text{in,s}}\}^{-1}\mathcal{A}^{-1}\{\vec{\gamma}_\text{s} + \vec{\gamma}_\text{schur}\}) \nonumber \\
	&= \left[\pi^{2M}\det(\cov_{\text{th,ee}}) \det(\cov_\text{schur} + \mathcal{A}\cov_{\text{in,s}} \mathcal{A}^\T)\right]^{-\frac{1}{2}} \nonumber \\
	&\qquad\times \exp(-[\vec{\gamma}_\text{s} + \vec{\gamma}_\text{schur}]^\T \{\cov_\text{schur} + \mathcal{A}\cov_{\text{in,s}}\mathcal{A}^\T\}^{-1}[\vec{\gamma}_\text{s} + \vec{\gamma}_\text{schur}])
,\end{align}
with a covariance matrix $\cov_\text{out} = \cov_\text{schur} + \mathcal{A}\cov_{\text{in,s}}\mathcal{A}^\T$ and a displacement in phase space of $\ev{\hat{\vec{\Gamma}}_\text{out}} = - \vec{\gamma}_\text{schur}$.
If the number of input modes exceeds the number of output modes, i.e. $N > M$, the output Wigner function simplifies to
\begin{align}
	W_\text{out}(\vec{\gamma}) &= \left[\pi^{2M} \det(\cov_B + \mathcal{A}\cov_{\text{in,s}} \mathcal{A}^\T)\right]^{-\frac{1}{2}} \exp(-\vec{\gamma}^\T \{\cov_B + \mathcal{A}\cov_{\text{in,s}}\mathcal{A}^\T\}^{-1}\vec{\gamma})
.\end{align}

\end{widetext}

\end{document}